%% file: main.tex
\documentclass[acmsmall]{acmart}

\acmJournal{PACMPL}
\acmVolume{1}
\acmNumber{POPL} 
\acmArticle{1}
\acmYear{2017}
\acmMonth{1}
\acmDOI{} 
\startPage{1}

\setcopyright{none}

\usepackage{amsmath}
\usepackage{amssymb}
\usepackage[plain,Figure]{algorithm}
\usepackage[noend]{algpseudocode}
\usepackage{graphicx}
\usepackage{stmaryrd}
\usepackage{mathtools}
\usepackage{xspace}
\usepackage{color}
\usepackage{comment}
\usepackage{multirow}
\usepackage{amsthm}
\usepackage{array}
\usepackage{booktabs}
\usepackage{pifont}
\usepackage{bbold}
\usepackage{wrapfig}
\usepackage{hyperref}
\usepackage{esvect}
\usepackage{soul}
\usepackage{caption}
\usepackage{booktabs}   
\usepackage{subcaption} 

\begin{document}

\title{Program Synthesis using Abstraction Refinement}



\author{Xinyu Wang}
\affiliation{
  \department{Computer Science}             
  \institution{University of Texas at Austin}            
  \city{Austin}
  \state{TX}
  \country{USA}                    
}
\email{xwang@cs.utexas.edu}          

\author{Isil Dillig}
\affiliation{
  \department{Computer Science}             
  \institution{University of Texas at Austin}           
  \city{Austin}
  \state{TX}
  \country{USA}                   
}
\email{isil@cs.utexas.edu}         

\author{Rishabh Singh}
\affiliation{
  \institution{Microsoft Research}           
  \city{Redmond}
  \state{WA}
  \country{USA}                   
}
\email{risin@microsoft.com}         

\input{macros}
\input{abstract}



\maketitle

\input{intro}

\input{prelim}

\input{abstractions}
\input{synthesis}

\input{example}

\input{impl}
\input{eval}

\input{related}

\input{conc}
\input{limitations}

\input{ack}


\input{appendix}

\bibliography{main}

\end{document}

%% file: macros.tex
\newcommand{\fta}{\mathcal{A}}
\newcommand{\ftastate}{q}
\newcommand{\ftastates}{Q}
\newcommand{\alphabet}{F}
\newcommand{\finalstates}{{\ftastates_f}}
\newcommand{\transitions}{\Delta}
\newcommand{\transition}{\delta}

\newcommand{\cval}{c}
\newcommand{\cvals}{{\vec{\cval}}}
\newcommand{\semantics}[1]{\llbracket{#1}\rrbracket}

\newcommand{\aval}{\varphi}
\newcommand{\universe}{\mathcal{U}}
\newcommand{\refines}{\sqsubseteq}
\newcommand{\pred}{p}
\newcommand{\preds}{\mathcal{P}}
\newcommand{\asemantics}[1]{{\semantics{#1}^\sharp}}

\newcommand{\abst}{\alpha^{\mathcal{P}}}
\newcommand{\abstpar}[1]{\alpha^{#1}}

\newcommand{\failure}{\text{null}}
\newcommand{\itp}{\mathcal{I}}
\newcommand{\prog}{\Pi}

\newcommand{\aeval}[1]{{\llbracket{#1}\rrbracket^\sharp_\preds}}

\newcommand{\grammar}{G}
\newcommand{\terminals}{T}
\newcommand{\nonterminals}{N}
\newcommand{\productions}{P}
\newcommand{\outputsymbol}{s_0}
\newcommand{\inputsymbol}{x}
\newcommand{\gsymbol}{s}
\newcommand{\terminal}{t}

\newcommand{\ex}{e}
\newcommand{\exs}{\vec{\ex}}
\newcommand{\inputex}{\ex_{\inp}}
\newcommand{\outputex}{\ex_{\out}}
\newcommand{\inputexs}{\vec{\inputex}}
\newcommand{\outputexs}{\vec{\outputex}}
\newcommand{\inp}{{\emph{in}}}
\newcommand{\out}{{\emph{out}}}

\renewcommand{\algref}[1]{Algorithm~\ref{alg:#1}}
\newcommand{\alglabel}[1]{\label{alg:#1}}
\newcommand{\figref}[1]{Fig.~\ref{fig:#1}}
\newcommand{\figlabel}[1]{\label{fig:#1}}
\newcommand{\exref}[1]{Example~\ref{exp:#1}}
\newcommand{\exlabel}[1]{\label{exp:#1}}
\newcommand{\defref}[1]{Definition~\ref{def:#1}}
\newcommand{\deflabel}[1]{\label{def:#1}}
\newcommand{\tabref}[1]{Table~\ref{def:#1}}
\newcommand{\tablabel}[1]{\label{def:#1}}
\newcommand{\lemmaref}[1]{Lemma~\ref{def:#1}}
\newcommand{\lemmalabel}[1]{\label{def:#1}}
\newcommand{\langu}{\mathcal{L}}

\newcommand{\todo}[1]{{\color{red}{#1}}}
\newcommand{\assign}{:=}

\renewcommand{\dots}{\cdots}
\renewcommand{\ldots}{\dots}

\newcommand{\irule}[2]{\mkern-2mu\displaystyle\frac{#1}{\vphantom{,}#2}\mkern-2mu}

\algnewcommand\Input{\textbf{input: }} 
\algnewcommand\Output{\textbf{output: }}
 
\newcommand\rott[2]{\rotatebox[origin=c]{#1}{#2}}
\newcommand\rot[1]{\rotatebox[origin=c]{90}{#1}}

\newcommand{\tool}{\textsc{Blaze}\xspace}
\newcommand{\prose}{Prose\xspace}
\newcommand{\enum}{ENUM-EQ\xspace}

%% file: abstract.tex
\begin{abstract}
We present a new approach to example-guided program synthesis based on \emph{counterexample-guided abstraction refinement}. Our method uses the abstract semantics of the underlying DSL to find a program $P$ whose \emph{abstract} behavior satisfies the examples. However, since program $P$ may be spurious with respect to the concrete semantics, our approach iteratively refines the abstraction until we either find a program that satisfies the examples or prove that no such DSL program exists. Because many programs have the same input-output behavior in terms of their \emph{abstract semantics}, this synthesis methodology significantly reduces the search space compared to existing techniques that use purely concrete  semantics.

While \emph{synthesis using abstraction refinement (SYNGAR)} could be implemented in different settings, we propose a refinement-based synthesis algorithm that uses \emph{abstract finite tree automata (AFTA)}. Our technique uses a coarse initial program abstraction to construct an initial AFTA, which is iteratively refined by constructing a \emph{proof of incorrectness} of any spurious program. In addition to ruling out the spurious program accepted by the previous AFTA, proofs of incorrectness are also useful for ruling out many other spurious programs.

We implement these ideas in a framework called \tool, which can be instantiated in different domains by providing a suitable DSL and its corresponding concrete and abstract semantics. We have used the \tool framework to build synthesizers for  string and matrix transformations, and we compare \tool with existing techniques. Our results for the string domain show that \tool compares favorably with FlashFill, a domain-specific synthesizer that is now deployed in Microsoft PowerShell. In the context of matrix manipulations, we compare \tool against \prose, a state-of-the-art general-purpose VSA-based synthesizer, and show that \tool results in a 90x speed-up over \prose. In both application domains, \tool also consistently improves upon the performance of two other existing techniques by at least an order of magnitude. 
\end{abstract}

%% file: intro.tex
\section{Introduction} 
In recent years, there has been significant interest in automatically synthesizing programs from input-output examples. Such programming-by-example (PBE) techniques have been successfully used to synthesize string and format transformations~\cite{flashfill,format} , automate data wrangling tasks~\cite{morpheus}, and synthesize programs that manipulate data structures~\cite{lambda2,hades,myth}. Due to its potential to automate many tasks encountered by end-users, programming-by-example has now become a burgeoning research area.

Because program synthesis is effectively a very difficult search problem, a key challenge  in this area is how to deal with the enormous size of the underlying search space. Even if we restrict ourselves to short programs of fixed length over a small domain-specific language, the synthesizer may still need to explore a colossal number of programs before it finds one that satisfies the  specification. In programming-by-example, a common search-space reduction technique exploits the observation that programs that yield the same \emph{concrete} output on the same input are indistinguishable with respect to the user-provided specification. Based on this observation, many techniques use a canonical representation of a large set of programs that have the same input-output behavior.  For instance,  enumeration-based techniques, such as Escher~\cite{escher} and Transit~\cite{transit},  discard programs that yield the same output as a previously explored program. Similarly, synthesis algorithms in the Flash* family~\cite{flashfill,flashmeta}, use a single node to represent all sub-programs that have the same input-output behavior. Thus, in all of these algorithms, the size of the search space is determined by the concrete output values produced by the DSL programs on the given inputs. 

In this paper, we aim to develop a more scalable general-purpose synthesis algorithm by using the \emph{abstract} semantics of DSL constructs rather than their concrete semantics. Building on the insight that we can reduce the size of the search space by exploiting commonalities in the input-output behavior of programs, our approach considers two programs to belong to the same equivalence class if they produce the same \emph{abstract} output on the same input. Starting  from the input example, our algorithm symbolically executes programs in the DSL using their \emph{abstract semantics} and merges any programs that have the same abstract output into the same equivalence class. The algorithm then looks for a program whose abstract behavior is consistent with the user-provided examples.  Because two programs that do not have the same input-output behavior in terms of their concrete semantics may have the same behavior in terms of their abstract semantics,  our approach has the potential to reduce the search space size in a more dramatic way.

\begin{figure}
\begin{center}
\includegraphics[scale=0.55]{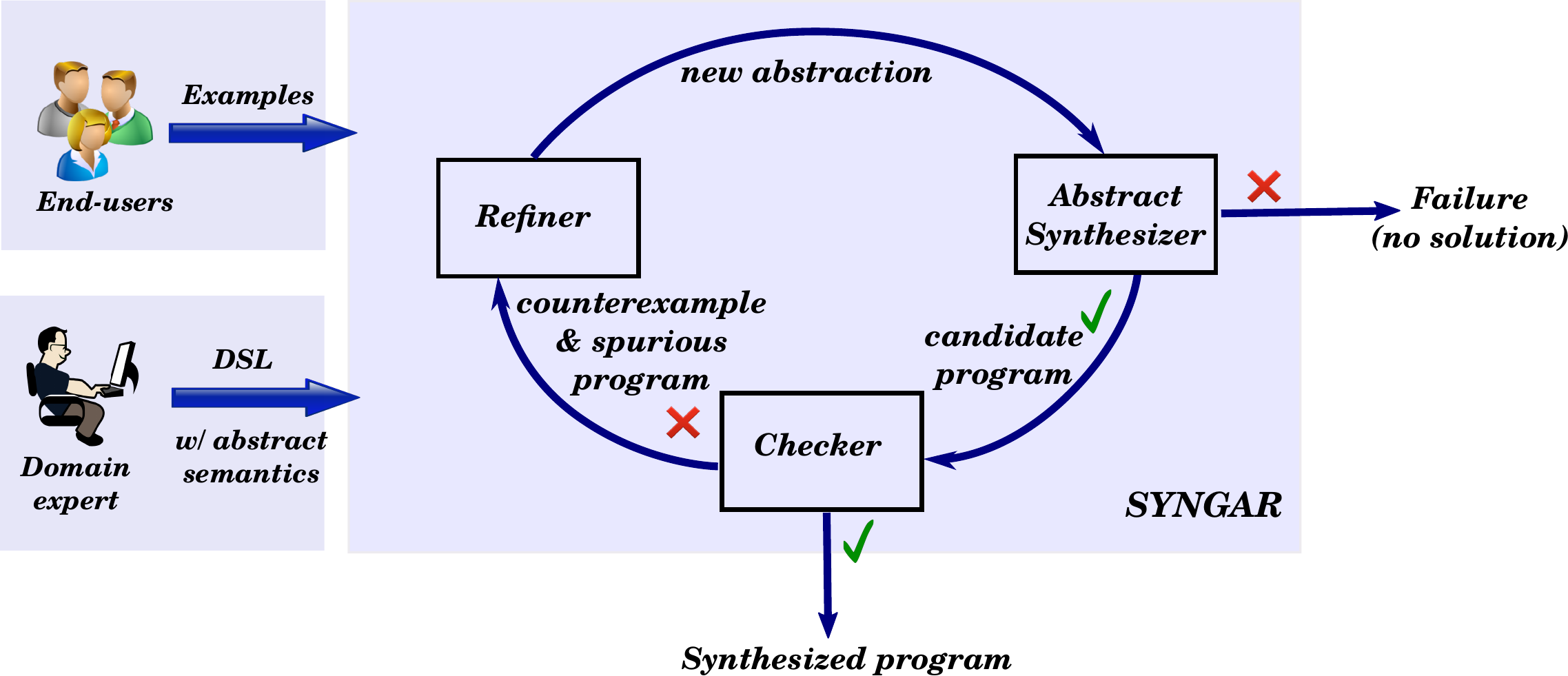}
\end{center}
\caption{Workflow illustrating \emph{\underline{\smash{Syn}}thesis usin\underline{\smash{g}} \underline{\smash{a}}bstraction \underline{\smash{r}}efinement (SYNGAR)}.  Since our approach is domain-agnostic, it is parametrized over a domain-specific language with both concrete and abstract semantics. From a user's perspective, the only input to the algorithm is a set of input-output examples.}\label{fig:refinement}
\vspace{-5pt}
\end{figure}

Of course, one obvious implication of such an abstraction-based approach is that the synthesized programs may now be \emph{spurious}: That is, a program that is consistent with the provided examples based on its abstract semantics may not actually satisfy the examples. Our synthesis algorithm iteratively eliminates such spurious programs by performing a form of \emph{counterexample-guided abstraction refinement}: Starting with a coarse initial abstraction, we first find a program $P$ that is consistent with the input-output examples with respect to its abstract semantics. If $P$ is also consistent with the examples using the concrete semantics, our algorithm returns $P$ as a solution. Otherwise, we refine the current abstraction, with the goal of ensuring that $P$ (and hopefully many other spurious programs) are no longer consistent with the specification using the new  abstraction. As shown in Figure~\ref{fig:refinement}, this refinement process continues until we either   find a program that satisfies the input-output examples, or prove that no such DSL program exists.

While the general idea of program synthesis using abstractions can be realized in different ways,
we develop this idea by generalizing a  recently-proposed synthesis algorithm that uses \emph{finite tree automata} (FTA)~\cite{dace}. The key idea underlying this  technique is to use   the \emph{concrete} semantics of the DSL to construct an FTA whose language is exactly the set of programs that are consistent with the input-output examples. While this approach  can, in principle,  be used to synthesize programs over any DSL, it suffers from the same scalability problems as other techniques that use concrete program semantics. 

In this paper, we introduce the notion of \emph{abstract finite tree automata (AFTA)}, which can be used to synthesize programs over the DSL's abstract semantics. Specifically, states in an AFTA  correspond to \emph{abstract} values and  transitions are constructed using the DSL's \emph{abstract} semantics. Any program accepted by the AFTA is consistent with the specification in the DSL's abstract semantics, but not necessarily in its concrete semantics. Given a spurious program $P$ accepted by the AFTA, 
our technique  automatically refines the current abstraction by constructing a so-called \emph{proof of incorrectness}.
Such a proof annotates the nodes of the abstract syntax tree representing $P$ with predicates that should be used in the new abstraction. 
The AFTA constructed  in the next iteration is guaranteed to reject $P$,   alongside many other spurious programs accepted by the previous AFTA. 


We have implemented our proposed idea in a synthesis framework called \tool, which can be instantiated in different domains by providing a suitable DSL with its corresponding  concrete and abstract semantics. As one application, we use \tool to automate string transformations from the SyGuS benchmarks~\cite{sygus} and empirically compare \tool against FlashFill, a synthesizer shipped with Microsoft PowerShell and that specifically targets string transformations~\cite{flashfill}. In another application, we have used \tool to automatically synthesize non-trivial matrix and tensor manipulations in MATLAB and compare \tool with \prose, a state-of-the-art synthesis tool based on version space algebra~\cite{flashmeta}. Our evaluation shows that \tool compares favorably with FlashFill in the string domain and that it outperforms \prose by 90x when synthesizing matrix transformations. 
We also compare \tool against  enumerative search techniques in the style of Escher and Transit~\cite{transit, escher} 
and show that  \tool results in at least an order of magnitude speedup for both application domains. Finally, we demonstrate the advantages of abstraction refinement by comparing \tool against a baseline synthesizer that constructs finite tree automata using the DSL's concrete semantics. 

\vspace{0.1in}\noindent
{\bf \emph{Contributions.}} To summarize, this paper makes the following key contributions:
\begin{itemize}
\item 
We propose a new synthesis methodology based on abstraction refinement. Our methodology reduces the size of the search space by using the abstract semantics of DSL constructs and automatically refines the abstraction whenever the synthesized program is spurious with respect to the input-output examples.
\item We introduce  \emph{abstract finite tree automata} and show how they can be used in program synthesis.
\item We describe a technique for automatically constructing a \emph{proof of incorrectness} of a spurious program and discuss how to use such proofs for abstraction refinement. 
\item  We develop a general synthesis framework called \tool, which can be instantiated in different domains by providing a suitable DSL with  concrete and abstract semantics.
\item We instantiate the \tool framework in two different domains involving string and matrix transformations. Our evaluation shows that \tool can synthesize non-trivial programs and that it results in significant improvement over existing techniques. Our evaluation also demonstrates the  benefits of  performing  abstraction refinement. 

\end{itemize}

\vspace{0.1in}\noindent
{\bf \emph{Organization.}} We first provide some background   on finite tree automata (FTA) and review a synthesis algorithm based on FTAs (Section~\ref{sec:background}). We then introduce \emph{abstract finite tree automata}  (Section~\ref{sec:afta}), describe our refinement-based synthesis algorithm (Section~\ref{sec:synthesis}), and then illustarate the technique using a concrete example (Section~\ref{sec:example}). The next section explains how to instantiate \tool in different domains and provides implementaion details. Finally, Section~\ref{sec:eval} presents  our experimental evaluation  and  Section~\ref{sec:related} discusses related work.


%% file: prelim.tex
\section{Preliminaries}\label{sec:background}

In this section, we give background on finite tree automata (FTA) and briefly review (a generalization of) an FTA-based synthesis algorithm proposed in previous work~\cite{dace}.

\subsection{Background on Finite Tree Automata}

A \emph{finite tree automaton} is a type of state machine that deals with tree-structured data. In particular, finite tree automata generalize standard finite automata by accepting trees rather than strings.

\begin{definition} {\bf (FTA)}
A (bottom-up) finite tree automaton (FTA) over alphabet $\alphabet$ is a tuple $\fta = (\ftastates, \alphabet, \finalstates, \transitions)$ where $\ftastates$ is a set of states, $\finalstates \subseteq \ftastates$ is a set of final states, and $\transitions$ is a set of transitions (rewrite rules)   of the form
$f(\ftastate_1, \dots, \ftastate_n) \rightarrow \ftastate$
where $\ftastate, \ftastate_1, \dots, \ftastate_n \in \ftastates$ and $f \in \alphabet$. 
\end{definition}

\begin{wrapfigure}{h}{0.48\linewidth}
\vspace{-0.16in}
\begin{center}
\includegraphics[scale=0.42]{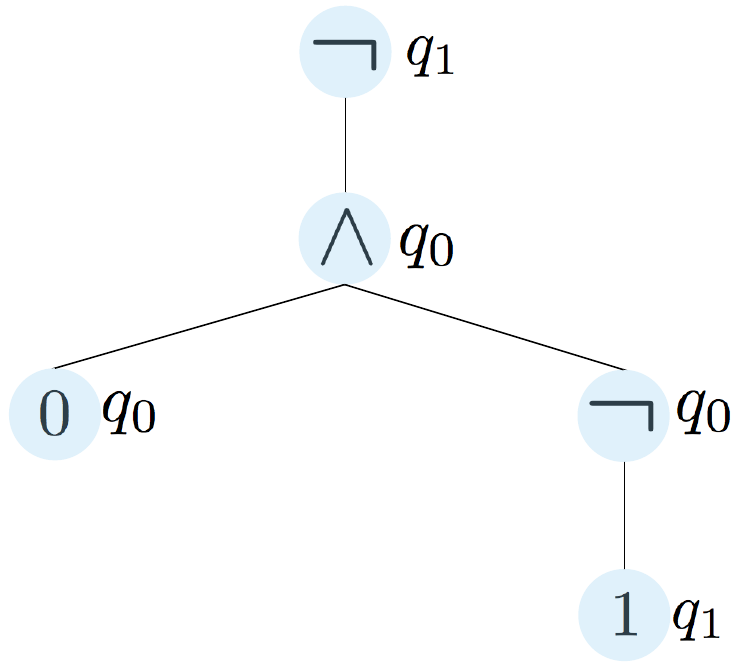}
\end{center}
\vspace{-0.1in}
\caption{Tree for $\neg(0 \wedge \neg 1)$, annotated with states.}
\figlabel{ftaexample}
\end{wrapfigure}

We assume that every symbol $f$ in alphabet $\alphabet$ has an arity (rank) associated with it, and we use the notation $\alphabet_k$ to denote the function symbols of arity $k$. We view ground terms over alphabet $\alphabet$ as trees such that a ground term $t$ is accepted by an FTA if we can rewrite $t$ to some state $\ftastate \in \finalstates$ using rules in $\transitions$. The language of an FTA $\fta$, denoted $\langu(\fta)$, corresponds to  the set of all ground terms accepted by $\fta$.

\begin{example} {\bf (FTA)}
Consider the tree automaton $\fta$ defined by states $\ftastates = \{ \ftastate_0, \ftastate_1 \}$, $\alphabet_0 = \{ 0, 1 \}$, $\alphabet_1 = \{ \neg \}$, $\alphabet_2 = \{ \wedge \}$, final states $\finalstates = \{ \ftastate_0 \}$, and the following transitions $\transitions$: 
\[
\small
\begin{array}{llll}
1 \rightarrow \ftastate_1 \ \ \ \ & 0 \rightarrow \ftastate_0 \ \ \ \ & \wedge(\ftastate_0, \ftastate_0) \rightarrow \ftastate_0 \ \ \ \ & \wedge(\ftastate_0, \ftastate_1) \rightarrow \ftastate_0 \\ 
\neg(\ftastate_0) \rightarrow \ftastate_1 \ \ \ \ & \neg(\ftastate_1) \rightarrow \ftastate_0 \ \ \ \ & \wedge(\ftastate_1, \ftastate_0) \rightarrow \ftastate_0 \ \ \ \ & \wedge(\ftastate_1, \ftastate_1) \rightarrow \ftastate_1 \\ 
\end{array}
\]
This tree automaton accepts those propositional logic formulas (without variables) that evaluate to \emph{false}. As an example, \figref{ftaexample} shows the tree for formula $\neg(0 \wedge \neg 1)$ where each sub-term is annotated with its state on the right. This formula is not accepted by the tree automaton $\fta$ because the  rules in $\transitions$ ``rewrite" the input to state $\ftastate_1$, which is not a final state. 
\end{example}

\subsection{Synthesis using Concrete Finite Tree Automata}\label{sec:cfta}

Since our approach builds on a prior synthesis technique that uses finite tree automata, we first review the key ideas underlying the work of \citet{dace}. However, since that work uses finite tree automata in the specific context of synthesizing data completion scripts, our formulation generalizes their approach to synthesis tasks over a broad class of DSLs.

Given a DSL and a set of input-output examples, the key idea is to  construct a finite tree automaton that represents the set of all DSL programs that are consistent with the input-output examples. The states of the FTA correspond to \emph{concrete} values, and the transitions are obtained using the \emph{concrete} semantics of the DSL constructs. We therefore refer to such tree automata  as \emph{concrete FTAs} (CFTA).

To understand the construction of CFTAs, suppose that we are given a set of input-output examples $
\exs$ and a context-free grammar $\grammar$ defining a DSL. We represent the input-output examples $\exs$  as a vector, where each element is of the form $\inputex \rightarrow \outputex$, and we write $\exs_\inp$ (resp. $\exs_\out$) to represent the input (resp. output) examples. Without loss of generality, we assume that programs take a single input $x$, as we can always represent multiple inputs as a list. Thus, the synthesized programs are always of the form $\lambda x. S$, and $S$ is defined by the grammar $\grammar =  (\terminals, \nonterminals, \productions, \outputsymbol)$ where:

\begin{itemize}
\item $T$ is a set of terminal symbols, including input variable $\inputsymbol$.  We refer to terminals other than $x$ as \emph{constants}, and use the notation $T_C$ to denote these constants.
\item $\nonterminals$ is a finite set of non-terminal symbols that represent sub-expressions in the DSL.
\item $\productions$ is a set of productions of the form $s \rightarrow f(s_1, \dots, s_n)$ where $f$ is a built-in DSL function and $s, s_1, \dots, s_n$ are symbols in the grammar. 
\item  $\outputsymbol \in \nonterminals$ is the topmost non-terminal (start symbol) in the grammar.
\end{itemize}

\begin{figure}[!t]
\footnotesize
\[
\begin{array}{cr}
\begin{array}{ccc}
\irule{
\vec{c} = \exs_\inp
}{
\ftastate_{\inputsymbol}^{\vec{c}} \in \ftastates 
} \ \ {\rm (Var)}
& \ \ \ \ \ \ \ 
\irule{
\begin{array}{c}
\terminal \in \terminals_C  \quad \vec{c} = \big[\semantics{\terminal}, \ldots, \semantics{\terminal}\big] \quad |\vec{c}| = |\exs|
\end{array}
}{
\ftastate_{t}^{\vec{c}} \in \ftastates 
} \ \ {\rm (Const)}
& \ \ \ \ \ \ \ 
\irule{
\ftastate_{\outputsymbol}^{\vec{c}} \in \ftastates  \quad \vec{c} = \exs_\out 
}{
\ftastate_{\outputsymbol}^{\vec{c}} \in \finalstates
} \ \  {\rm (Final)}
\end{array}
\\ \\
\irule{
\begin{array}{c}
(s \rightarrow f(s_1, \ldots, s_n)) \in P \quad q_{s_1}^{\vec{c_1}} \in \ftastates, \ldots, q_{s_n}^{\vec{c_n}} \in \ftastates 
\quad
c_{j} = \semantics{f(c_{1j}, \ldots, c_{nj})} \quad \vec{c} = [c_{1}, \ldots, c_{|\exs|}]
\quad 
\end{array}
}{
\ftastate_{s}^{\vec{c}} \in \ftastates,  \quad \big( f(q_{s_1}^{\vec{c_1}}, \ldots, q_{s_n}^{\vec{c_n}}) \rightarrow \ftastate_{s}^{\vec{c}} \big) \in \transitions
}\ \ {\rm (Prod)}
\end{array}
\]
\vspace{-10pt}
\caption{Rules for constructing CFTA $\fta = (\ftastates, \alphabet, \finalstates, \transitions)$ given examples $\exs$ and grammar $\grammar =   (\terminals, \nonterminals, \productions, \outputsymbol)$.}
\figlabel{concreterules}
\end{figure}

We can construct the CFTA for examples $\exs$ and grammar $\grammar$ using the rules shown in~\figref{concreterules}.  First, the alphabet of the CFTA consists of the built-in functions (operators) in the DSL. The states in the CFTA are of the form $\ftastate_s^{\vec{c}}$, where $s$ is a symbol (terminal or non-terminal) in the grammar and $\vec{c}$ is a vector of concrete values. Intuitively, the existence of a state $\ftastate_s^{\vec{c}}$ indicates that symbol $s$ can take concrete values $\vec{c}$ for input examples $\exs_\inp$. Similarly, the existence of a transition $f(q_{s_1}^{\vec{c_1}}, \ldots, q_{s_n}^{\vec{c_n}}) \rightarrow \ftastate_s^{\vec{c}}$ means that applying function $f$ on the concrete values $c_{1j}, \ldots, c_{nj}$ yields $c_j$. Hence, as mentioned earlier, transitions of the CFTA are constructed using the concrete semantics of the DSL constructs.

We now briefly explain the rules from \figref{concreterules} in more detail. The first rule, labeled Var, states that $q_x^{\vec{c}}$ is a state whenever $x$ is the input variable and $\vec{c}$ is the input examples. 
The second rule, labeled Const, adds a state $\ftastate_\terminal^{[\semantics{t}, \ldots, \semantics{t}]}$ for each constant $\terminal$ in the grammar.  The next rule, called Final, indicates that $q_{\outputsymbol}^{\vec{c}}$ is a final state whenever $\outputsymbol$ is the start symbol in the grammar and $\vec{c}$ is the output examples. 
The last rule, labeled Prod, generates new CFTA states and transitions for each production $s \rightarrow f(s_1, \ldots, s_n)$. Essentially, this rule states that, if symbol $s_i$ can take value $\vec{c}_i$ (i.e., there exists a state $q_{s_i}^{\vec{c}_i}$) and executing $f$ on $c_{1j}, \ldots, c_{nj}$ yields value  $c_j$, then we also have a state $\ftastate_s^{\vec{c}}$ in the CFTA and a transition $f(\ftastate_{s_1}^{\vec{c}_1} \ldots, \ftastate_{s_n}^{\vec{c}_n}) \rightarrow \ftastate_s^{\vec{c}}$.

It can be shown that the language of the CFTA constructed from \figref{concreterules} is exactly the set of abstract syntax trees (ASTs) of DSL programs that are consistent with the input-output examples.~\footnote{The proof can be found in the extended version of this paper~\cite{syngarextended}.} Hence, once we construct such a CFTA, the synthesis task boils down to finding an AST that is accepted by the automaton. However, since there are typically many ASTs accepted by the CFTA, one can use heuristics to identify the ``best" program that satisfies the input-output examples.

\vspace{0.1in}\noindent
{\bf \emph{Remark.}}  In general, the tree automata constructed using the rules from \figref{concreterules} may have infinitely many states. As standard in synthesis literature~\cite{flashmeta, solarthesis}, we therefore assume that the size of programs under consideration should be less than a given bound.  In terms of the CFTA construction, this means we only add a state $\ftastate_s^\cvals$ if the size of the smallest tree accepted by the automaton $ (\ftastates, \alphabet, \{ \ftastate_s^\cvals\}, \transitions)$ is lower than the threshold.

\begin{example} 
To see how to construct CFTAs, let us consider the following very simple toy DSL, which only contains two constants and allows addition and multiplication by constants:
\[
\small 
\begin{array}{lll}
n & \assign & id(x) \ | \ n + t \ | \ n \times t; \\ 
t & \assign & 2 \ | \ 3 ; \\ 
\end{array}
\]
Here, \emph{id} is just the identity function. The CFTA representing the set of all DSL programs with at most two $+$ or $\times$ operators for the input-output example $1 \rightarrow 9$ is shown in \figref{cfta}. For readability, we use circles to represent states of the form $q_n^c$, diamonds to represent $q_x^c$ and squares to represent $q_t^c$, and the number labeling the node shows the value of $c$. There is a state $q_x^1$ since the value of $x$ is $1$ in the provided  example (Var rule). We construct transitions using the concrete semantics of the DSL constructs (Prod rule). For instance, there is a transition $\emph{id}(q_x^1) \rightarrow q_n^1$ because  $\emph{id}(1)$ yields value $1$ for symbol $n$. Similarly, there is a transition $+(q_n^1, q_t^2) \rightarrow q_n^3$ since the result of adding $1$ and $2$ is $3$. The only accepting state is $q_n^9$ since the start symbol in the grammar is $n$ and the output has value $9$ for the given example. This CFTA accepts two programs, namely $(id(x) + 2 ) \times 3$ and $(id(x) \times 3) \times 3$. Observe that these are the only two programs with at most two $+$ or $\times$ operators in the DSL that are consistent with the example $1 \rightarrow 9$.
\exlabel{toydsl}
\end{example}

\begin{figure}[!t]
\centering
\includegraphics[scale=0.3]{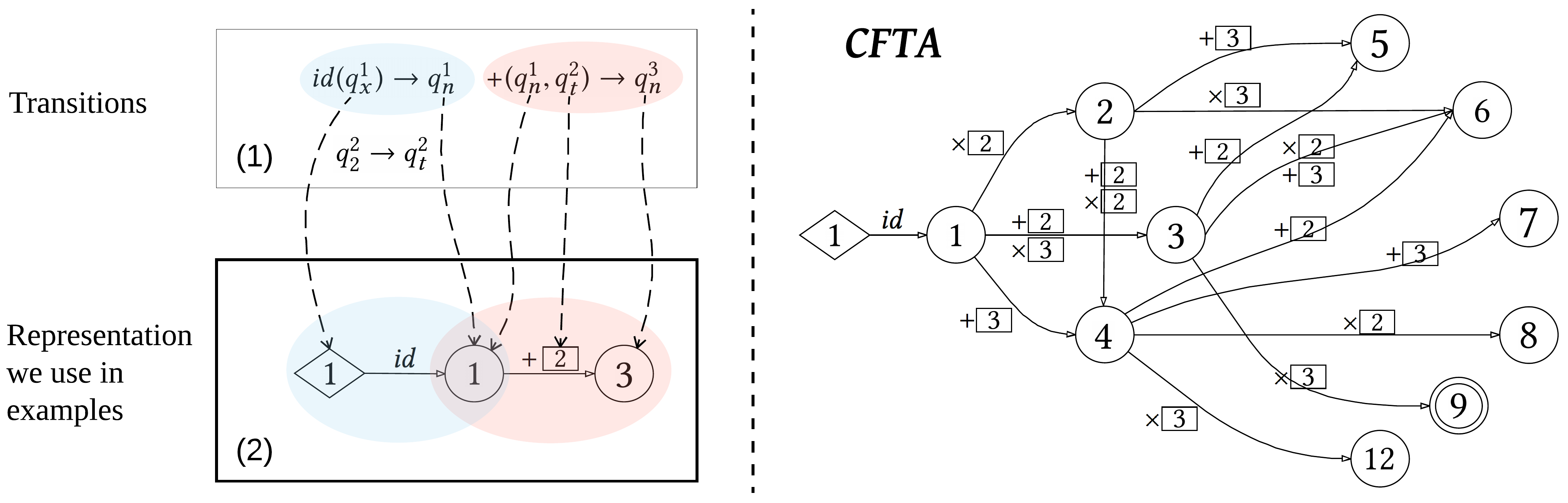}
\vspace{-10pt}
\caption{The CFTA constructed for \exref{toydsl}. We visualize the CFTA as a graph where nodes are labeled with concrete values for symbols. Edges correspond to transitions and are labeled with the operator (i.e., $+$ or $\times$) followed by the constant operand (i.e., $2$ or $3$). For example, for the upper two transitions shown in (1) on the left, the graphical representation is shown in (2). Moreover, to make our representation easier to view, we do not include transitions that involve nullary functions in the graph. For instance, the transition $q_2^2 \rightarrow \ftastate_{t}^{2}$ in (1) is not included in (2). A transition of the form $f(\ftastate_{n}^{c_1}, \ftastate_{t}^{c_2}) \rightarrow \ftastate_{n}^{c_3}$ is represented by an edge from a node labeled $c_1$ to another node labeled $c_3$, and the edge is labeled by $f$ followed by $c_2$. For instance, the transition $+(\ftastate_{n}^1, \ftastate_t^2) \rightarrow \ftastate_n^3$ in (1) is represented by an edge from $1$ to $3$ with label $+2$ in (2).}
\figlabel{cfta}
\end{figure}

%% file: abstractions.tex
\section{Abstract Finite Tree Automata}\label{sec:afta}


In this section, we introduce \emph{abstract finite tree automata (AFTA)}, which  form the basis of the synthesis algorithm that we will present in Section~\ref{sec:synthesis}. 
However, since our approach performs \emph{predicate abstraction} over the concrete values of  grammar symbols, we first start by reviewing some requirements on the underlying abstract domain.

\subsection{Abstractions}

In the previous section, we saw that CFTAs associate a \emph{concrete value} for each grammar symbol by executing the concrete semantics of the DSL on the user-provided inputs. To construct abstract FTAs, we will instead associate an \emph{abstract value} with each symbol. In the rest of the paper, we assume that abstract values are represented as \emph{conjunctions} of predicates of the form $f(s)\ op\ c$, where $s$ is a symbol in the grammar defining the DSL, $f$ is a function, and $c$ is a constant. For example, if symbol $s$ represents an array, then predicate $len(s) > 0$ may indicate that the array is non-empty.  Similarly, if $s$ is a matrix, then $\emph{rows}(s) = 4$ could indicate that $s$ contains exactly 4 rows.

\vspace{0.1in}\noindent
{\bf \emph{Universe of predicates.}} As mentioned earlier, our approach is parametrized over a DSL constructed by a \emph{domain expert}.  We will assume that the domain expert also specifies a suitable universe $\universe$ of  predicates that may appear in the abstractions used in our synthesis algorithm. In particular, given a family  of functions $\mathcal{F}$, a set of operators $\mathcal{O}$, and a set of constants $\mathcal{C}$ specified by the domain expert, the universe $\universe$ includes any predicate $f(s)\ op\ c$ where  $f \in \mathcal{F}$, $op \in \mathcal{O}$,  $c \in \mathcal{C}$, and $\gsymbol$ is a grammar symbol. To ensure the completeness of our approach, we require that $\mathcal{F}$ always contains the identity function, $\mathcal{O}$ includes equality, and $\mathcal{C}$ includes all concrete values that symbols in the grammar can take. As we will see, this requirement  ensures that every CFTA can be expressed as an AFTA over our predicate abstraction. We also assume that the universe of predicates includes \emph{true} and \emph{false}. In the remainder of this paper, we use the notation $\universe$ to denote the universe of all possible predicates that can be used in our algorithm.

\vspace{0.1in}\noindent
{\bf \emph{Notation.}} Given two abstract values $\aval_1$ and $\aval_2$,  we write $\aval_1 \sqsubseteq \aval_2$ iff the formula $\aval_1 \Rightarrow \aval_2$ is logically valid. As standard in abstract interpretation~\cite{cousot77}, we write $\gamma(\aval)$ to denote the set of concrete values represented by abstract value $\aval$. Given predicates $\mathcal{P} = \{p_1, \ldots, p_n\} \subseteq \universe$ and a formula (abstract value) $\aval$ over universe $\universe$, we  write $\abst(\aval)$ to denote the strongest conjunction of predicates $p_i \in \mathcal{P}$ that is logically implied by $\aval$.
Finally, given a vector of abstract values $\vec{\aval} = [ \aval_1, \ldots, \aval_n ]$, we  write  $\abst(\vec{\aval})$ to mean $\vec{\aval}'$ where $\aval'_i = \abst(\aval_i)$.

\vspace{0.1in}\noindent
{\bf \emph{Abstract semantics.}}
In addition to specifying the universe of predicates, we assume that the domain expert also specifies the abstract semantics of each DSL construct by providing symbolic post-conditions over the universe of predicates $\universe$. We represent the abstract semantics for a production $s \rightarrow f(s_1, \dots, s_n)$ using the notation $\asemantics{f(\aval_1, \dots, \aval_n)}$. That is, given abstract values $\aval_1, \dots, \aval_n$ for the argument symbols $s_1, \dots, s_n$, the abstract transformer $\asemantics{f(\aval_1, \dots, \aval_n)}$ returns an abstract value $\aval$ for $s$. We require that the abstract transformers are \emph{sound}, i.e.:
\[
\emph{If} \ \ \asemantics{f(\aval_1, \dots, \aval_n)} = \aval \ \emph{and} \   c_1 \in \gamma(\aval_1), \dots,  c_n \in \gamma(\aval_n), \ \emph{then} \ \semantics{f(c_1, \dots, c_n)} \in \gamma(\aval)
\]

However, in general, we do not require the abstract transformers to be \emph{precise}. That is, if we have $\asemantics{f(\aval_1, \dots, \aval_n)} = \aval$ and $S$ is the set containing $\semantics{f(c_1, \ldots, c_n)}$ for every $c_i \in \gamma(\aval_i)$, then
it is possible that $\aval \sqsupseteq \abstpar{\universe}(S)$.
In other words, we allow each abstract transformer to produce an abstract value that is weaker (coarser) than the value produced by the most precise transformer over the given abstract domain.
We do not require the abstract semantics to be precise because it may be cumbersome to define the most precise abstract transformer for some DSL constructs.
On the other hand, we require an abstract transformer $\asemantics{f(\aval_1, \dots, \aval_n)}$ where each $\aval_i$ is of the form $s_i = c_i$ to be precise. Note that this can be easily implemented using the concrete semantics:
\[
\asemantics{f(s_1 = c_1, \dots, s_n = c_n)} = (s = \semantics{f(c_1, \dots, c_n)})
\]


{
\begin{example} 
Consider the same DSL that we used in \exref{toydsl} and suppose the universe $\universe$  includes \emph{true},  all predicates of the form $ x = c$, $t=c$, and $n=c$ where $c$ is an integer, and predicates $0 <n \leq 4, 0 < n \leq 8$. 
Then, the abstract semantics can be defined as follows:
\[
\footnotesize 
\begin{array}{ll}
\asemantics{id ( x = c )} \assign (n = c) \\ \\ 
\asemantics{(n = c_1) + (t = c_2)} \assign (n = (c_1 + c_2) ) & 
\asemantics{(n = c_1) \times (t = c_2)} \assign (n = c_1 c_2) \\
\asemantics{(0 < n \leq 4) + (t = c)} \assign \left\{
\begin{array}{ll}
0 < n \leq 4 & c=0 \\
0 < n \leq 8 & 0 < c \leq 4 \\
\emph{true} & \text{otherwise}
\end{array}
\right.
& 
\asemantics{(0 < n \leq 4) \times (t = c)} \assign \left\{
\begin{array}{ll}
0 < n \leq 4 & c=1  \\ 
0 <n \leq 8 & c=2 \\
\emph{true} & \text{otherwise}
\end{array}
\right.
\\ 
\asemantics{(0 < n \leq 8) + (t = c)} \assign \left\{
\begin{array}{ll}
0 < n \leq 8 & c = 0 \\ 
\emph{true} & \text{otherwise} \\ 
\end{array}
\right.
& 
\asemantics{(0 < n \leq 8) \times (t = c)} \assign \left\{ 
\begin{array}{ll}
0 < n \leq 8 & \text{c = 1} \\ 
\emph{true} & \text{otherwise} \\
\end{array}
\right.
\\ \\ 
\asemantics{ \big(\bigwedge_i{\pred_i} \big) \diamond \big( \bigwedge_j{\pred_j} \big)} \assign \bigsqcap_i \bigsqcap_j { \asemantics{\pred_i \diamond \pred_j}} &  \diamond \in \{ +, \times \} \\
\end{array}
\]
In addition,  the abstract transformer returns \emph{true} if any of its arguments is \emph{true}. 
\exlabel{atransformers}
\end{example}
}

\subsection{Abstract Finite Tree Automata}

As mentioned earlier, abstract finite tree automata (AFTA) generalize concrete FTAs by associating abstract -- rather than concrete -- values with each symbol in the grammar. Because an abstract value  can represent \emph{many} different concrete values, multiple states in a CFTA might correspond to a \emph{single} state in the AFTA. Therefore, AFTAs typically have far fewer states than their corresponding CFTAs, allowing us to construct and analyze them much more efficiently than CFTAs.

\begin{figure}[!t]
\footnotesize
\[
\begin{array}{cr}
\begin{array}{cc}
\irule{
\vec{\aval} = \abst \big( \big[ x = \exs_{\inp, 1}, \dots, x = \exs_{\inp, | \exs |} \big] \big) 
}{
\ftastate_{\inputsymbol}^{\vec{\aval}} \in \ftastates 
} \ \ {\rm (Var)}
& \ \ \ \ \ 
\irule{
\begin{array}{c}
\terminal \in \terminals_C  \quad \vec{\aval} = \abst \big( \big[ \terminal = \semantics{\terminal}, \ldots, \terminal = \semantics{\terminal} \big] \big) \quad |\vec{\aval}| = |\exs|
\end{array}
}{
\ftastate_{\terminal}^{\vec{\aval}} \in \ftastates 
} \ \ {\rm (Const)}
\end{array}
\\ \\

\irule{
\ftastate_{\outputsymbol}^{\vec{\aval}} \in \ftastates \quad 
\forall j \in [1, |\exs_\out|]. \ (\outputsymbol = \ex_{\emph{out},j}) \refines  \aval_j
}{
\ftastate_{\outputsymbol}^{\vec{\aval}} \in \finalstates
} \ \  {\rm (Final)}
\\ \\
\hspace*{-5pt}
\irule{
\begin{array}{c}
(s \rightarrow f(s_1, \ldots, s_n)) \in P \quad q_{s_1}^{\vec{\aval_1}} \in \ftastates, \ldots, q_{s_n}^{\vec{\aval_n}} \in \ftastates 
\quad
\aval_{j} = \abst\big(\asemantics{f(\aval_{1j}, \ldots, \aval_{nj})}\big)  \quad \vec{\aval} = [\aval_{1}, \ldots, \aval_{|\exs|}]
\quad 
\end{array}
}{
\ftastate_{s}^{\vec{\aval}} \in \ftastates,  \quad \big( f(q_{s_1}^{\vec{\aval_1}}, \ldots, q_{s_n}^{\vec{\aval_n}}) \rightarrow \ftastate_{s}^{\vec{\aval}} \big) \in \transitions
}\ \ {\rm (Prod)}
\end{array}
\]
\caption{Rules for constructing AFTA $\fta = (\ftastates, \alphabet, \finalstates, \transitions)$ given examples $\exs$, grammar $\grammar =   (\terminals, \nonterminals, \productions, \outputsymbol)$ and a set of predicates $\mathcal{P} \subseteq \universe$.}
\figlabel{abstractrules}
\end{figure}

States in an AFTA are of the form $\ftastate_{s}^{\vec{\aval}}$ where $s$ is a symbol in the grammar and $\vec{\aval}$ is a vector of abstract values. If there is a transition $f(\ftastate_{s_1}^{\vec{\aval_1}}, \dots, \ftastate_{s_n}^{\vec{\aval_n}}) \rightarrow \ftastate_{s}^{\vec{\aval}}$ in the AFTA, it is always the case that $\asemantics{f(\aval_{1j}, \dots, \aval_{nj})} \refines \aval_{j}$.  Since our abstract transformers are sound, this means that $\aval_j$ overapproximates the result of running $f$ on the concrete values represented by $\aval_{1j}, \ldots, \aval_{nj}$.

Let us now consider the AFTA construction rules shown in \figref{abstractrules}. Similar to CFTAs, our construction requires the set of input-output examples $\exs$ as well as the grammar $\grammar =  (\terminals, \nonterminals, \productions, \outputsymbol)$ defining the DSL. In addition, the AFTA construction requires the abstract semantics of the DSL constructs (i.e., $\asemantics{f(\ldots)}$) as well as a set of predicates $\preds \subseteq \universe$ over which we construct our abstraction. 

The first two rules from \figref{abstractrules} are very similar to their counterparts from the CFTA construction rules: According to the Var rule, the states $\ftastates$ of the AFTA include a state $\ftastate_x^{\vec{\aval}}$ where $x$ is the input variable and $\vec{\aval}$ is the abstraction  of the input examples $\exs_\inp$ with respect to predicates $\mathcal{P}$. Similarly, the Const rules states that $\ftastate_\terminal^{\vec{\aval}} \in \ftastates$ whenever $\terminal$ is a constant (terminal) in the grammar and $\vec{\aval}$ is the abstraction of $[t = \semantics{t}, \ldots, t = \semantics{t}]$ with respect to predicates $\mathcal{P}$.
The next rule, labeled  Final in \figref{abstractrules}, defines the final states of the AFTA. Assuming  the start symbol in the grammar is $\outputsymbol$, then $\ftastate_{\outputsymbol}^{\vec{\aval}}$ is a final state whenever the concretization of $\vec{\aval}$ includes the output examples.

The last rule, labeled Prod, deals with grammar productions of the form $s \rightarrow f(s_1, \ldots, s_n)$. Suppose that the AFTA contains states $\ftastate_{s_1}^{\vec{\aval_1}}, \ldots, \ftastate_{s_n}^{\vec{\aval_n}}$, which, intuitively, means that symbols $s_1, \ldots, s_n$ can take abstract values $\vec{\aval}_1, \ldots, \vec{\aval}_n$. In the Prod rule, we first ``run" the abstract transformer for $f$ on abstract values $\aval_{1j}, \ldots, \aval_{nj}$ to obtain an abstract value $\asemantics{f(\aval_{1j}, \ldots, \aval_{nj})}$ over the universe $\universe$. However, since the set of predicates $\mathcal{P}$ may be a strict subset of the universe $\universe$, $\asemantics{f(\aval_{1j}, \ldots, \aval_{nj})}$ may not be a valid abstract value with respect to predicates $\mathcal{P}$. Hence, we apply the abstraction function $\abst$ to $\asemantics{f(\aval_{1j}, \ldots, \aval_{nj})}$ to find the strongest conjunction $\aval_j$ of predicates over $\mathcal{P}$ that overapproximates $\asemantics{f(\aval_{1j}, \ldots, \aval_{nj})}$. Since symbol $s$ in the grammar can take abstract value $\vec{\aval}$, we add the state $\ftastate_s^{\vec{\aval}}$ to the AFTA, as well as the transition  $ f(q_{s_1}^{\vec{\aval_1}}, \ldots, q_{s_n}^{\vec{\aval_n}}) \rightarrow \ftastate_{s}^{\vec{\aval}}$.

\begin{wrapfigure}{h}{0.3\linewidth}
\centering
\vspace*{-7pt}
\includegraphics[scale=0.6]{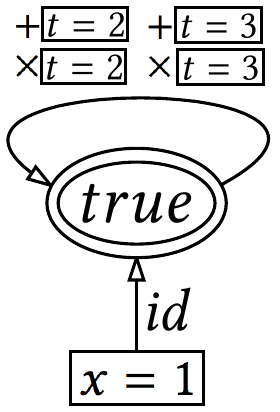}
\caption{AFTA in \exref{afta}.}
\figlabel{afta}
\end{wrapfigure}

\begin{example} 
Consider the same DSL that we used in \exref{toydsl} as well as the universe and abstract transformers  given in \exref{atransformers}. Now, let us consider the set of predicates  $\preds = \{ \emph{true}, t=2, t=3, x=c \}$ where $c$ stands for any integer value. \figref{afta} shows the AFTA constructed for the input-output example  $1 \rightarrow 9$ over predicates $\preds$. Since the abstraction of $x=1$ over $\preds$ is $x=1$, the AFTA includes a state $q_x^{x=1}$, shown simply as $x=1$. Since $\preds$ only has $\emph{true}$ for symbol $n$, the AFTA contains a transition $\emph{id}(q_x^{x=1}) \rightarrow q_n^\emph{true}$, where $q_n^\emph{true}$ is abbreviated as \emph{true} in \figref{afta}. The AFTA also includes transitions $+(q_n^\emph{true}, t = c) \rightarrow q_n^\emph{true}$ and $\times(q_n^\emph{true}, t = c) \rightarrow q_n^\emph{true}$ for $c \in \{2, 3\}$. Observe that $q_n^\emph{true}$  is the only final state since $n$ is the start symbol and the concretization of \emph{true} includes $9$ (the output example). Thus, the language of this AFTA includes all programs that start with $\emph{id}(x)$.
\exlabel{afta}
\end{example}

\begin{theorem}{\bf (Soundness of AFTA)}
Let $\fta$ be the AFTA constructed for examples $\exs$ and grammar $\grammar$ using the abstraction defined by finite set of predicates $\mathcal{P}$ (including $\emph{true}$). If $\prog$ is a program that is consistent with examples $\exs$, then $\prog$ is accepted by $\fta$. 
\end{theorem}
\begin{proof}
Please find the proof in the extended version of this paper~\cite{syngarextended}.
\end{proof}

%% file: synthesis.tex
\section{Synthesis using Abstraction Refinement}\label{sec:synthesis}

We now turn our attention to the top-level synthesis algorithm using abstraction refinement. The key idea underlying our technique is to construct an abstract FTA using a coarse initial abstraction. We then iteratively refine this abstraction and its corresponding AFTA until we either find a program that is consistent with the input-output examples or prove that there is no DSL program that satisfies them. In the remainder of this section, we first explain the top-level synthesis algorithm and then describe the auxiliary procedures in later subsections.

\subsection{Top-level Synthesis Algorithm}\label{sec:topsynthesis}

The high-level structure of our refinement-based synthesis algorithm is shown in \figref{synthesisalgorithm}. The {\sc Learn} procedure from \figref{synthesisalgorithm} takes as input a set of examples $\exs$, a grammar $G$ defining the DSL, an initial set of predicates $\mathcal{P}$, and the universe of all possible predicates $\universe$. We implicitly assume that we also have access to  the concrete and abstract semantics of the DSL. Also, it is worth noting that the initial set of predicates $\mathcal{P}$ is optional. In cases where the domain expert does not specify  $\mathcal{P}$, the initial abstraction includes $\emph{true}$, predicates of the form $x = c$ where $c$ is any value that has the same type as $x$, and predicates of the form $t = \semantics{t}$ where $t$ is a constant (terminal) in the grammar.

Our synthesis algorithm consists of a refinement loop (lines 2--9), in which we alternate between AFTA construction, counterexample generation, and predicate learning.  In each iteration of the refinement loop, we first construct an AFTA using the current set of predicates $\preds$ (line 3). If the language of the AFTA is empty, we have a proof that there is no DSL program that satisfies the input-output examples; hence, the  algorithm returns \emph{null} in this case (line 4). Otherwise, we use a heuristic \emph{ranking algorithm} to choose a ``best''  program $\prog$ that is accepted by the current AFTA $\fta$ (line 5). In the remainder of this section, we assume that programs are represented as abstract syntax trees where each  node is labeled with the corresponding DSL construct.  We do not fix a particular ranking algorithm for {\sc Rank}, so the synthesizer is free to choose between any number of different ranking heuristics as long as {\sc Rank} returns a program that has the lowest cost with respect to a deterministic cost metric.

Once we find a program $\prog$ accepted by the current AFTA, we run it on the input examples $\exs_\inp$ (line 6). If the result matches the expected outputs $\exs_\out$, we  return $\prog$ as a solution of the synthesis algorithm. Otherwise, we refine the current abstraction so that the spurious program $\prog$ is no longer accepted by the refined AFTA. Towards this goal, we  find a single input-output example $e$ that is inconsistent with program $\prog$ (line 7), i.e., a \emph{counterexample}, and then construct a \emph{proof of incorrectness} $\itp$ of $\prog$ with respect to the counterexample $e$ (line 8). In particular,  $\itp$  is a mapping from the AST nodes in $\prog$ to abstract values over universe $\universe$ and provides a proof (over the abstract semantics) that program $\prog$ is inconsistent with example $e$. More formally, a proof of incorrectness $\itp$ must satisfy the following definition:


\begin{definition}{\bf (Proof of Incorrectness)}\label{def:itp}
Let $\prog$ be the AST of a program that does not satisfy example $\ex$. Then, a \emph{proof of incorrectness} of $\prog$ with respect to $\ex$ has the following properties:
\begin{enumerate}
\item If $v$ is a leaf node of $\prog$ with label $t$, then $( t = \semantics{t}\ex_\inp ) \refines \itp(v)$. 
\item If $v$ is an internal node with label $f$ and children $v_1, \ldots, v_n$, then: 
\[ \asemantics{f ( \itp(v_1), \ldots, \itp(v_n) )} \refines \itp(v) \]
\item If  $\itp$ maps the root node of $\prog$  to  $\aval$, then $e_\out \not \in \gamma(\aval)$.

\end{enumerate}
\end{definition}

Here, the first two properties state that $\itp$ constitutes a proof (with respect to the abstract semantics) that executing $\prog$ on input $\ex_\inp$ yields an output that satisfies $\itp(\emph{root}(\prog))$. The third property states that $\itp$ proves that $\prog$ is spurious, since $\ex_\out$ does not satisfy $\itp(\emph{root}(\prog))$. The following theorem states that a proof of incorrectness of a spurious program always exists. 

\begin{theorem}{\bf (Existence of Proof)}
Given a spurious program $\prog$ that does not satisfy example $\ex$, we can always find a proof of incorrectness of $\prog$ satisfying the properties from Definition~\ref{def:itp}.
\end{theorem}
\begin{proof}
Please find the proof in the extended version of this paper~\cite{syngarextended}.
\end{proof}

\begin{figure}[!t]
\small  
\begin{algorithm}[H]
\begin{algorithmic}[1]
\Procedure{Learn}{$\exs, \grammar, \mathcal{P}, \universe$} 
\vspace{3pt}
\Statex \Input{Input-output examples $\exs$,  context-free grammar $\grammar$,  initial predicates $\mathcal{P}$, and universe $\universe$.} 
\Statex \Output{A program consistent with the examples.} 
\vspace{3pt}
\While{true} \Comment{Refinement loop.}
\State $\fta \assign \textsc{ConstructAFTA}(\exs, \grammar,  \mathcal{P})$; 
\If{$\mathcal{L}(\fta) = \emptyset$} \Return{$\emph{null}$;} \EndIf
\State $\prog \assign \textsc{Rank}(\fta)$;
\If{$\semantics{\prog} \exs_\inp = \exs_\out$} \Return $\prog$; \EndIf
\State $\ex \assign \textsc{FindCounterexample}(\prog, \exs)$; \Comment{$\ex \in \exs \text{ and } \semantics{\prog} \ex_\inp \neq \ex_\out$.}
\State $\itp \assign \textsc{ConstructProof}(\prog, \ex, \preds, \universe)$;
\State $\mathcal{P} \assign \mathcal{P} \bigcup \textsc{ExtractPredicates}(\itp)$;
\EndWhile
\EndProcedure
\end{algorithmic}
\end{algorithm}
\vspace{-30pt}
\caption{The top-level structure of our synthesis algorithm using abstraction refinement.}
\figlabel{synthesisalgorithm}
\end{figure}

Our synthesis algorithm uses such a proof of incorrectness $\itp$ to refine the current abstraction. In particular, the predicates that we use in the next iteration include all predicates that appear in $\itp$ in addition to the old set of predicates $\mathcal{P}$. Furthermore, as stated by the following theorem, the AFTA constructed in the next iteration is guaranteed to \emph{not accept} the spurious program $\prog$ from the current iteration. 

\begin{theorem}{\bf (Progress)}
Let $\fta_i$ be the AFTA constructed during the i'th iteration of the {\sc Learn} algorithm from \figref{synthesisalgorithm}, and let $\prog_i$ be a spurious program returned by {\sc Rank}, i.e., $\prog_i$ is accepted by $\fta_i$ and does not satisfy input-output examples $\ex$. Then, we have $\prog_i \not \in \mathcal{L}(\fta_{i+1})$ and $\mathcal{L}(\fta_{i+1}) \subset \mathcal{L}(\fta_{i})$.
\end{theorem}
\begin{proof}
Please find the proof in the extended version of this paper~\cite{syngarextended}.
\end{proof}

\begin{example} 
Consider the AFTA constructed in \exref{afta}, and suppose the program returned by {\sc Rank} is $id(x)$. Since this program is  inconsistent with the input-output example $1 \rightarrow 9$, our algorithm constructs the proof of incorrectness shown in \figref{proof}. In particular, the proof labels the root node of the AST with the new abstract value $0 < n \leq 8$, which establishes that $id(x)$ is spurious because $9 \not \in \gamma(0<n \leq 8)$. In the next iteration, we add $0 < n \leq 8$ to our set of predicates $\preds$ and construct the new AFTA shown in \figref{proof}. Observe that the spurious program $\emph{id}(x)$ is no longer accepted by the refined AFTA.
\end{example}

\begin{wrapfigure}{r}{0.45\linewidth}
\centering
\vspace*{-20pt}
\includegraphics[scale=0.55]{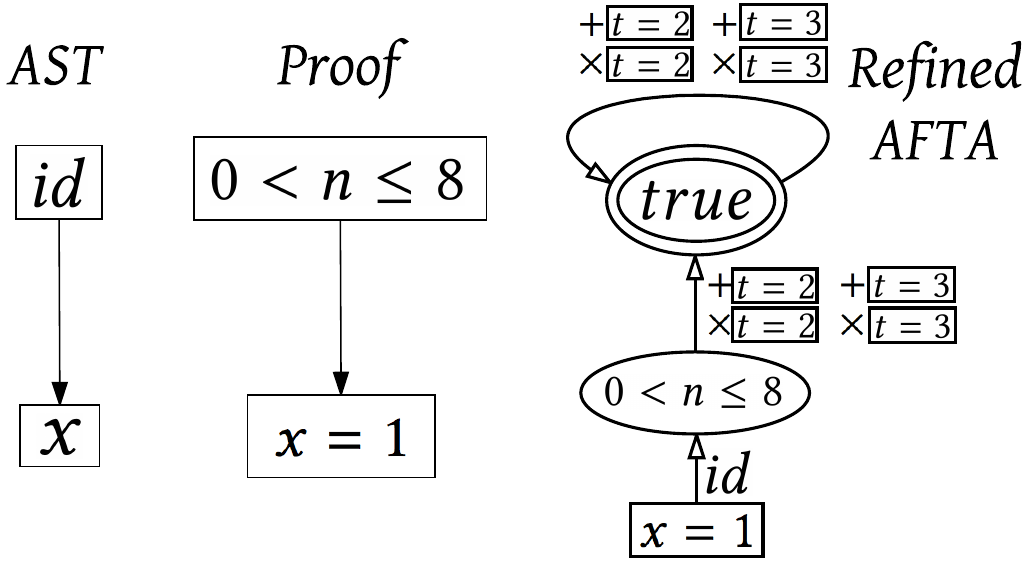}
\caption{Proof of incorrectness for \exref{afta}.}
\figlabel{proof}
\end{wrapfigure}

\begin{theorem}{\bf (Soundness and Completeness)}
If there exists a DSL program that satisfies the input-output examples $\exs$, then the {\sc Learn} procedure from \figref{synthesisalgorithm} will return a program $\prog$ such that $\semantics{\prog} \exs_\inp = \exs_\out$.
\end{theorem}
\begin{proof}
Please find the proof in the extended version of this paper~\cite{syngarextended}.
\end{proof}

\subsection{Constructing Proofs of Incorrectness}

In the previous subsection, we saw how proofs of incorrectness are used to rule out spurious programs from the search space (i.e., language of the AFTA). We now discuss how to automatically construct such proofs given a spurious program.

Our algorithm for constructing a proof of incorrectness is shown in \figref{itp}. The {\sc ConstructProof} procedure takes as input a spurious program $\prog$ represented as an AST with vertices $V$ and an input-output example $e$ such that $\semantics{\prog} \ex_\inp \neq \ex_\out$. The procedure also requires the current abstraction defined by predicates $\preds$ as well as the universe of all predicates $\universe$. The output of this procedure is a mapping from the vertices $V$ of $\prog$ to new abstract values proving that $\prog$ is inconsistent with $\ex$.

At a high level, the {\sc ConstructProof} procedure processes the AST top-down, starting at the root node $r$. Specifically, we first find an annotation $\itp(r)$ for the root node such that $\ex_\out \not \in \gamma(\itp(r))$. In other words, the annotation $\itp(r)$ is sufficient for showing that $\prog$ is spurious (property (3) from Definition~\ref{def:itp}). After we find an annotation for the root node $r$ (lines 2--4), we add $r$ to \emph{worklist} and find suitable annotations for the children of all  nodes in the worklist. In particular, the loop in lines 6--15 ensures that $\itp$ satisfies properties (1) and (2) from Definition~\ref{def:itp}.

\begin{figure}[!t]
\small
\begin{algorithm}[H]
\begin{algorithmic}[1]
\Procedure{ConstructProof}{$\prog, \ex, \preds, \universe$} 
\vspace{3pt}
\Statex \Input{A spurious program $\prog$ represented as an AST with vertices $V$.}
\Statex  \Input{A counterexample $\ex$ such that $\semantics{\prog} \ex_{\inp} \neq \ex_{\out}$.}
\Statex \Input{Current set of predicates $\preds$ and the universe of predicates $\universe$.} 
\Statex \Output{A proof $\itp$ of incorrectness of $\prog$ represented as mapping from $V$ to abstract values over $\universe$.} 
\vspace{3pt}

\Statex \Comment{Find annotation $\itp(r)$ for root $r$ such that $\ex_\out \not \in \gamma(\itp(r))$.}
\State $\aval$ \assign \textsf{EvalAbstract}$(\prog, \ex_\inp, \preds)$; 
\State $\psi$ \assign \textsf{StrengthenRoot}$\big (\outputsymbol =  \semantics{\prog} \ex_\inp, \aval, \outputsymbol \neq \ex_\out, \universe \big)$; 
\State $\itp(\textsf{root}(\prog)) \assign \aval \land \psi$; 

\Statex \Comment{Process all nodes other than root.}
\State $\emph{worklist} \assign \big\{ \textsf{root}(\prog) \big\}$;
\While{$\emph{worklist} \neq \emptyset$}

\Statex \Comment{Find annotation $\itp(v_i)$ for each $v_i$  s.t $\asemantics{f(\itp(v_1), \ldots, \itp(v_n))} \refines \itp(\emph{cur})$.}
\State $\emph{cur} \assign \emph{worklist}.\textsf{remove}()$;
\State $\vec{\prog} \assign \textsf{ChildrenASTs}(\emph{cur})$;
\vspace{3pt}
\State $\vec{\phi} \assign \big[s_i = c_i \ \big | \ c_i = \semantics{\prog_i} \ex_\inp , i \in [1, |\vec{\prog}|], s_i = \textsf{Symbol}(\prog_i) \ \big]$;

\State $\vec{\aval} \assign \big[\aval_i \ \big | \ \aval_i = \textsf{EvalAbstract}(\prog_i, \ex_\inp, \preds), i \in [1, |\vec{\prog}|] \ \big]$;
\State $\vec{\psi} \assign \textsf{StrengthenChildren}\big (\vec{\phi}, \vec{\aval}, \itp(\emph{cur}), \universe, \textsf{label}(cur) \big)$;
\For{$i = 1, \dots, |\vec{\prog}|$}
\State $\itp(\textsf{root}(\prog_i)) \assign \aval_i \land \psi_i$;
\If{$\neg$\textsf{IsLeaf}(\textsf{root}($\prog_i$))}
\State {\ \emph{worklist}.\textsf{add}(\textsf{root}($\prog_i$))};
\EndIf
\EndFor
\EndWhile
\vspace{3pt}
\State \Return $\itp$;
\EndProcedure
\end{algorithmic}
\end{algorithm}
\vspace{-25pt}
\caption{Algorithm for constructing proof of incorrectness of $\prog$ with respect to example $\ex$. In the algorithm, \textsf{ChildrenASTs}($v$) returns the sub-ASTs rooted at the children of $v$. The function $\textsf{Symbol}(\prog)$ yields the grammar symbol for the root node of $\prog$.}
\figlabel{itp}
\end{figure}

\begin{figure}
\vspace{-10pt}
\small
\[
\begin{array}{lll}
\textsf{EvalAbstract}(\textsf{Leaf(x)}, \ex_\inp, \preds) & = \alpha^\preds(x = \ex_\inp)  \\
\textsf{EvalAbstract}(\textsf{Leaf(t)}, \ex_\inp, \preds) & = \alpha^\preds \big( t = \semantics{t} \big) \\
\textsf{EvalAbstract}(\textsf{Node}(f, \vec{\prog}), \ex_\inp, \preds) & = \alpha^\preds \Big(\asemantics{f\big (\textsf{EvalAbstract}(\prog_1, \ex_\inp, \preds), \ldots, \textsf{EvalAbstract}\big(\prog_{|\vec{\prog}|}, \ex_\inp, \preds)\big)} \Big)  \\
\end{array}
\]
\vspace{-10pt}
\caption{Definition of auxiliary \textsf{EvalAbstract} procedure used in {\sc ComputeProof} algorithm from \figref{itp}. 
\textsf{Node}($f, \vec{\prog}$) represents an internal node with label $f$ and subtrees $\vec{\prog}$.}
\figlabel{abstracteval}
\end{figure}

Let us now consider the {\sc ConstructProof} procedure in more detail. To find the annotation for the root node $r$, we first compute $r$'s  abstract value in the domain defined by  predicates $\preds$. Towards this goal, we use a procedure called \textsf{EvalAbstract}, shown in \figref{abstracteval}, which symbolically executes $\prog$ on $\ex_\inp$ using the abstract transformers (over $\preds$). The return value $\aval$ of $\textsf{EvalAbstract}$ at line 2 has the property that $\ex_\out \in \gamma(\aval)$, since the AFTA constructed using predicates $\preds$ yields the spurious program $\prog$. We then try to \emph{strengthen} $\aval$ using a new formula $\psi$ over predicates $\universe$ such that the following properties hold:

\begin{enumerate}
\item $(\outputsymbol = \semantics{\prog} \ex_\inp) \Rightarrow \psi$  where $\outputsymbol$ is the start symbol of the grammar,
\item $\aval \land \psi \Rightarrow  ( \outputsymbol \neq \ex_\out )$.
\end{enumerate}
Here, the first property says that the output of $\prog$ on input $\ex_\inp$ should satisfy $\psi$; otherwise $\psi$ would not be a correct strengthening. The second property says that $\psi$, together with the previous abstract value $\aval$, should be strong enough to show that $\prog$ is  \emph{inconsistent} with the input-output example $\ex$.  

While any strengthening $\psi$ that satisfies these two properties will be sufficient to prove that $\prog$ is spurious, we would ideally want our strengthening to rule out many other spurious programs. 
For this reason, we want  $\psi$ to be as general (i.e., logically weak) as possible. Intuitively, the more general the proof, the more spurious programs it can likely prove  incorrect. For example, while a predicate such as $\outputsymbol = \semantics{\prog}\ex_\inp$ can prove that $\prog$ is incorrect, it only proves the spuriousness of  programs that produce the same concrete output as $\prog$ on $\ex_\inp$. On the other hand, a more general predicate that is logically weaker than $\outputsymbol = \semantics{\prog}\ex_\inp$ can potentially prove the spuriousness of other programs that may not necessarily return the same concrete output as $\prog$ on $\ex_\inp$.

\begin{figure}[!t]
\footnotesize 
\begin{algorithm}[H]
\begin{algorithmic}[1]
\Procedure{StrengthenRoot}{$\pred_+,\pred_-,  \aval,  \universe$} 
\vspace{3pt}
\Statex \Input{Predicates $\pred_+$ and $\pred_-$, formula $\aval$, and universe $\universe$.}
\Statex \Output{Formula $\psi^*$ such that $\pred_+ \Rightarrow (\aval \land \psi^*) \Rightarrow \pred_-$.}
\vspace{3pt}
\State $\Phi \assign \big\{ \pred \in \universe\ \big| \ \pred_+ \Rightarrow \pred \big\}$; \ \  $\Psi \assign \Phi$; \Comment{Construct  universe of relevant predicates.}
\vspace{3pt}
\For{$i = 1, \dots, k$}  \Comment{Generate all possible conjunctions up to length $k$.}
\State $\Psi \assign \Psi \bigcup \big\{ \psi \wedge \pred \ \big| \  \psi \in \Psi, \pred \in \Phi \big\}$; 
\EndFor
\State $\psi^* \assign \pred_+$; \Comment{Find most general formula with desired property.}
\For{$\psi \in \Psi$}
\If{$\psi^* \Rightarrow \psi \ { \rm  and} \  (\aval \wedge \psi) \Rightarrow \pred_-$}
$\psi^* \assign \psi$; 
\EndIf
\EndFor
\vspace{3pt}
\State \Return $\psi^*$;
\EndProcedure
\end{algorithmic}
\end{algorithm}
\vspace{-30pt}
\caption{Algorithm for finding a strengthening for the root.}
\figlabel{strengthenroot}
\end{figure}

\begin{figure}[!t]
\vspace{-20pt}
\footnotesize 
\begin{algorithm}[H]
\begin{algorithmic}[1]
\Procedure{StrengthenChildren}{$\vec{\phi},  \vec{\aval}, \aval_p, \universe, f$} 
\vspace{3pt}
\Statex \Input{Predicates $\vec{\phi}$, formulas $\vec{\aval}$, formula  $\aval_p$,  and universe $\universe$.}
\Statex \Output{Formulas $\vec{\psi}^*$ such that $\forall i \in [1, |\vec{\psi}^*|]. \   \phi_i \Rightarrow \psi_i^*$ and
$\asemantics{f(\aval_1 \land \psi_1^* \ldots, \aval_n \land \psi_n^*) } \Rightarrow \aval_p$.}
\vspace{3pt}
\State $\vec{\Phi} \assign \big[ \Phi_i \ \big| \ \Phi_i = \big\{ \pred \in \universe \ \big| \ \phi_i \Rightarrow \pred \big\}\big]$; \ \ 
$\vec{\Psi} \assign \vec{\Phi}$ \Comment{Construct universe of relevant predicates.}
\vspace{3pt}
\For{$i = 1, \dots, k$} \Comment{Generate all possible conjunctions up to length $k$.}
\For{$j=1, \ldots, |\vec{\Psi}|$}
\State $\Psi_j \assign \Psi_j \bigcup \big\{ \psi \wedge \pred \ \big| \  \psi \in \Psi_j, \pred \in \Phi_j \big\}$
\EndFor
\EndFor
\State $\vec{\psi}^* \assign \vec{\phi}$; \Comment{Find most general formula with desired property.}
\ForAll{$\vec{\psi}$ \ {\rm where} $\psi_i \in \Psi_i$ }
\If{$\forall i \in [1, |\vec{\phi}|]. \ \psi_i^* \Rightarrow \psi_i$ and $\asemantics{f(\aval_{1} \wedge \psi_1, \dots, \aval_n \wedge \psi_n)} \Rightarrow \aval_p$}
$\vec{\psi}^* \assign \vec{\psi}$;
\EndIf
\EndFor
\vspace{3pt}
\State \Return $\vec{\psi}^*$;
\EndProcedure
\end{algorithmic}
\end{algorithm}
\vspace{-30pt}
\caption{Algorithm for finding a strengthening for nodes other than the root.}
\figlabel{strengthenchildren}
\end{figure}

To find such a suitable strengthening $\psi$, our algorithm makes use of a procedure called \textsf{StrengthenRoot}, described in \figref{strengthenroot}. In a nutshell, this procedure returns the most general conjunctive formula $\psi$ using at most $k$ predicates in $\universe$ such that the above two properties are satisfied.  Since $\psi$, together with the old abstract value $\aval$, proves the spuriousness of $\prog$, our proof $\itp$ maps the root node to the new strengthened abstract value $\aval \land \psi$ (line 4 of {\sc ConstructProof}). 

The loop in lines 5--15 of {\sc ConstructProof} finds annotations for all nodes other than the root node.  Any AST node $\emph{cur}$ that has been removed from the worklist at line 7 has the property that $\emph{cur}$ is in the domain of $\itp$ (i.e., we have already found an annotation for $\emph{cur}$). Now, our goal is to find a suitable annotation for $\emph{cur}$'s children such that $\itp$ satisfies properties (1) and (2) from Definition~\ref{def:itp}. To find the annotation for each child $v_i$ of $\emph{cur}$, we first compute the concrete  and abstract values ($\phi_i$ and $\aval_i$ from lines 9--10) associated with each $v_i$.  We then invoke the \textsf{StrengthenChildren} procedure, shown in \figref{strengthenchildren}, to find a strengthening $\vec{\psi}$ such that:

\begin{enumerate}
\item  $\forall i \in [1, |\vec{\psi}|]. \   \phi_i \Rightarrow \psi_i$
\item $\asemantics{f(\aval_1 \land \psi_1, \ldots, \aval_n \land \psi_n)} \Rightarrow \itp(\emph{cur})$
\end{enumerate}

Here, the first  property ensures that $\itp$ satisfies property (1) from Definition~\ref{def:itp}. In other words, the first condition says that our strengthening overapproximates the concrete output of subprogram $\prog_i$ rooted at $v_i$ on input $\ex_\inp$. The second condition enforces property (2) from Definition~\ref{def:itp}. In particular, it says that the annotation for the parent node is provable from the annotations of the children using the abstract semantics of the DSL constructs.

In addition to satisfying these  afore-mentioned properties, the strengthening $\vec{\psi}$ returned by {\sc StrengthenChildren} has some useful generality guarantees. In particular, the return value of the function is pareto-optimal in the sense that we cannot obtain a valid strengthening $\vec{\psi}'$ (with a fixed number of conjuncts) by weakening any of the $\psi_i$'s in $\vec{\psi}$. As mentioned earlier, finding such \emph{maximally general} annotations is useful because it allows our synthesis procedure to  rule out many spurious programs in addition to the specific one returned by the ranking algorithm.

\begin{example}
To better understand how we construct proofs of incorrectness, consider the AFTA shown in \figref{constructproof}(1). Suppose that the ranking algorithm returns the program $id(x) + 2$, which is clearly spurious with respect to the input-output example $1 \rightarrow 9$. \figref{constructproof}(2)-(4) show the AST for the program $\emph{id}(x) + 2$ as well as the old abstract and concrete values  for each AST node. Note that the abstract values from \figref{constructproof}(3) correspond to the results of  \textsf{EvalAbstract} in the {\sc ConstructProof} algorithm from \figref{itp}. Our proof construction algorithm starts by strengthening the root node $v_1$ of the AST. Since $\semantics{\prog} \ex_{\inp}$ is $3$, the first argument of the \textsf{StrengthenRoot} procedure is provided as $n=3$. Since the output value in the example is $9$, the second argument is $n \neq 9$. Now, we invoke the \textsf{StrengthenRoot} procedure to find a formula $\psi$ such that $n = 3 \Rightarrow (\emph{true} \wedge \psi) \Rightarrow n \neq 9$ holds. The most general conjunctive formula over $\universe$ that has this property is $0 < n \leq 8$; hence, we obtain the annotation $\itp(v_1) = 0 < n \leq 8$ for the root node of the AST. The {\sc ConstructProof} algorithm now ``recurses down" to the children of $v_1$ to find suitable annotations for $v_2$ and $v_3$. When processing $v_1$ inside the while loop in \figref{itp}, we have $\vec{\phi} = [n=1, t=2]$ since $1, 2$ correspond to the concrete values for $v_2, v_3$. Similarly, we have $\vec{\aval} = [0 < n \leq 8, t=2]$ for the abstract values for $v_2$ and $v_3$. We now invoke \textsf{StrenthenChildren} to find a $\vec{\psi} = [\psi_1, \psi_2]$ such that:
\[
\begin{array}{c}
n=1 \Rightarrow \psi_1 \quad  \quad t=2 \Rightarrow \psi_2 \\
\asemantics{+(0 < n \leq 8 \land \psi_1, \ t =2 \land \psi_2)} \Rightarrow 0 < n \leq 8
\end{array}
\]
In this case, \textsf{StrengthenChildren} yields the solution $\psi_1 = 0 < n \leq 4$ and $\psi_2 =  \emph{true}$. Thus, we have $\itp(v_2) = 0 < n \leq 4$ and $\itp(v_3) = (t = 2)$. The final proof of incorrectness for this example is shown in \figref{constructproof}(5).
\exlabel{constructproof}
\end{example}

\begin{figure}[!t]
\centering
\includegraphics[scale=0.46]{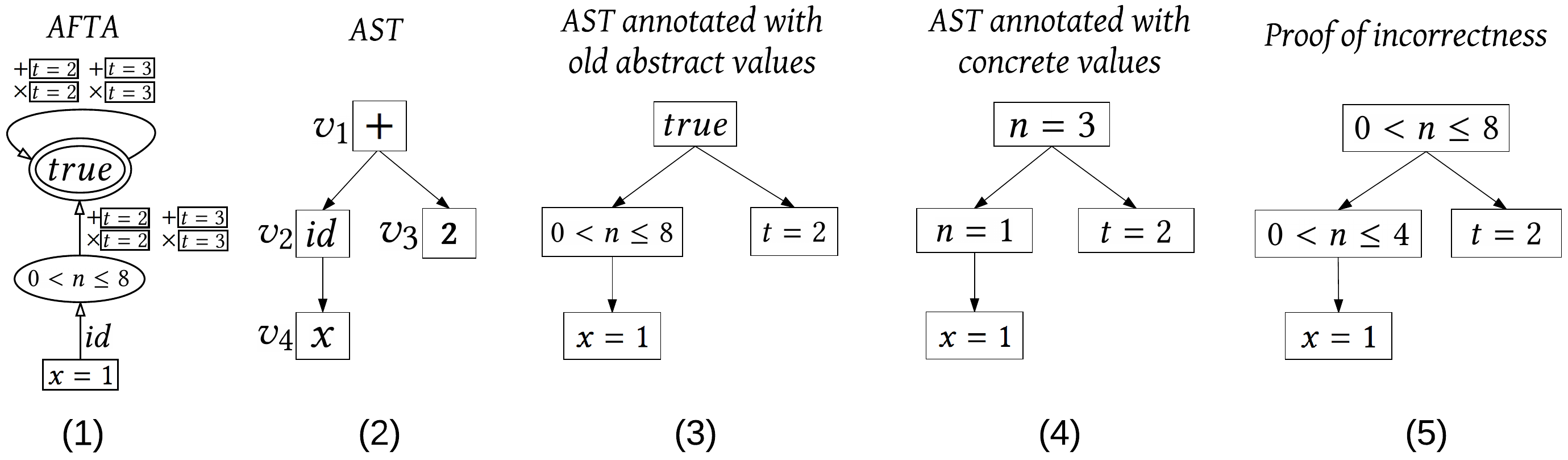}
\vspace{-5pt}
\caption{Illustration of the proof construction process for \exref{constructproof}.}
\figlabel{constructproof}
\vspace{-10pt}
\end{figure}

\vspace{-10pt}
\begin{theorem}{\bf (Correctness of Proof)}
The mapping $\itp$ returned by the {\sc ConstructProof} procedure satisfies the properties from Definition~\ref{def:itp}.
\end{theorem}
\begin{proof}
Please find the proof in the extended version of this paper~\cite{syngarextended}.
\end{proof}

\noindent 
{\bf \emph{Complexity analysis.}}
The complexity of our synthesis algorithm is mainly determined by the number of iterations, and the complexity of FTA construction, ranking and proof construction. In particular, the FTA can be constructed in time $\mathcal{O}(m)$ where $m$ is the size of the resulting FTA\footnote{FTA size is defined to be $\sum_{\transition \in \transitions}{|\transition|}$ where $|\transition| = n+1$ for a transition $\transition$ of the form $f(q_1, \dots, q_n) \rightarrow q$.} (before any pruning). The complexity of performing ranking over an FTA depends on the ranking heuristic. For the one used in our implementation (see Section~\ref{sec:implrank}), the time complexity is $\mathcal{O}(m \cdot \log d)$ where $m$ is the FTA size and $d$ is the number of states in the FTA. The complexity of proof construction for an AST is $\mathcal{O}(l \cdot p)$ where $l$ is the number of nodes in the AST and $p$ is the number of conjunctions under consideration. Therefore, the complexity of our synthesis algorithm is given by $\mathcal{O}(t \cdot (l \cdot p + m \cdot \log d))$ where $t$ is the number of iterations of the abstraction refinement process.

%% file: example.tex
\section{A Working Example}\label{sec:example}

In the previous sections, we illustrated various aspects of our synthesis algorithm using the DSL from \exref{toydsl} on the input-output example $1 \mapsto 9$. We now walk through the entire algorithm and show how it synthesizes the desired program $(\emph{id}(x)+2) \times 3$. We use the abstract semantics and universe of predicates $\universe$ given in \exref{atransformers}, and we use the initial set of predicates $\preds$ given in \exref{afta}. We will assume that the ranking algorithm always favors smaller programs over larger ones. In the case of a tie, the ranking algorithm favors programs that use $+$ and those that use smaller constants.

\begin{figure*}
\begin{minipage}{\linewidth}
\includegraphics[scale=0.48]{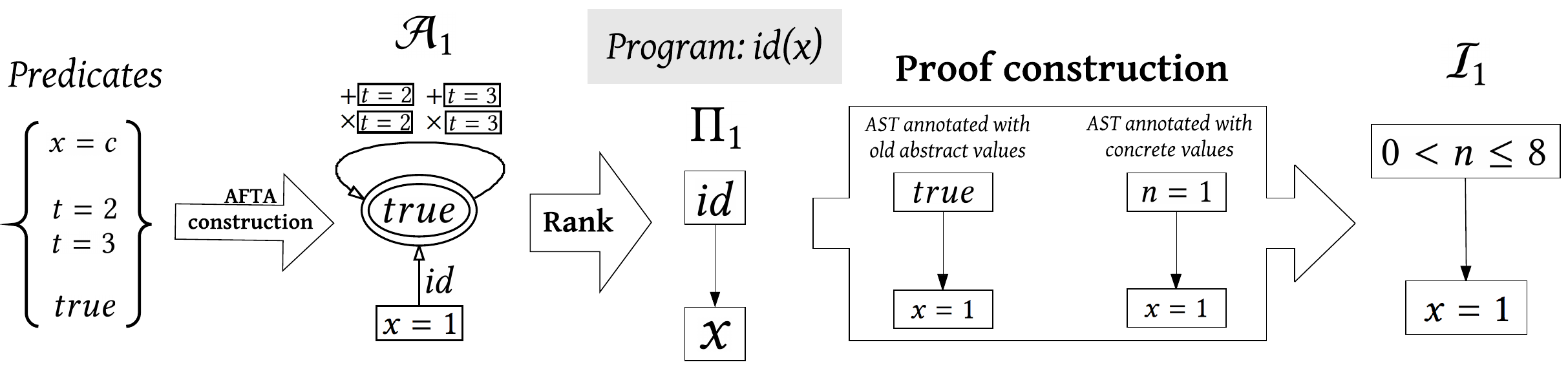}
\vspace{-5pt}
\caption*{Iteration 1: The constructed AFTA is $\fta_1$, {\sc Rank} returns $\prog_1$, $\prog_1$ is spurious, and the proof of incorrectness is $\itp_1$.}
\vspace{10pt}
\end{minipage}
\begin{minipage}{\linewidth}
\includegraphics[scale=0.43]{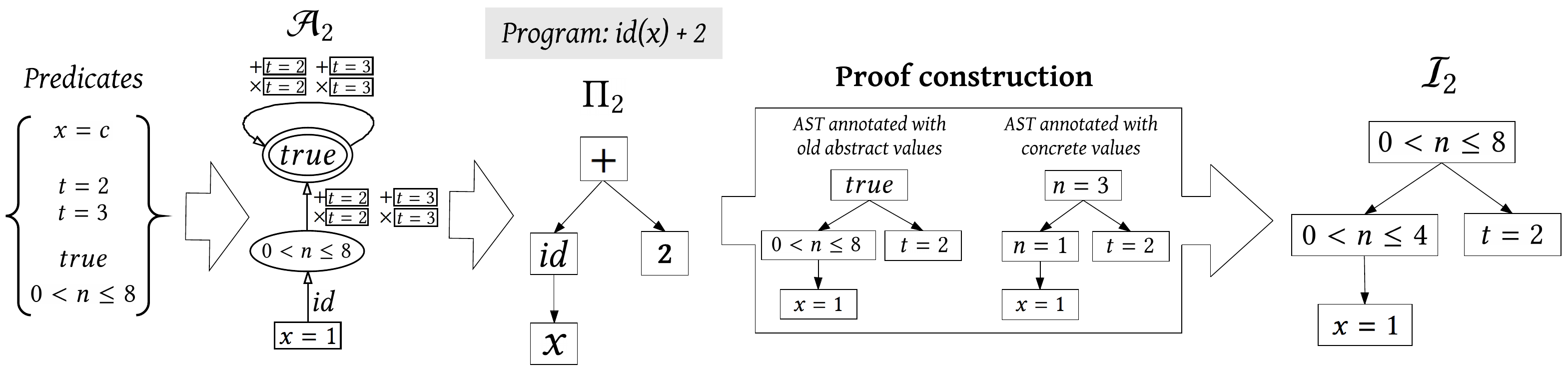}
\vspace{-5pt}
\caption*{Iteration 2: The constructed AFTA is $\fta_2$, {\sc Rank} returns $\prog_2$, $\prog_2$ is spurious, and the proof of incorrectness is $\itp_2$.}
\vspace{5pt}
\end{minipage}
\begin{minipage}{\linewidth}
\includegraphics[scale=0.43]{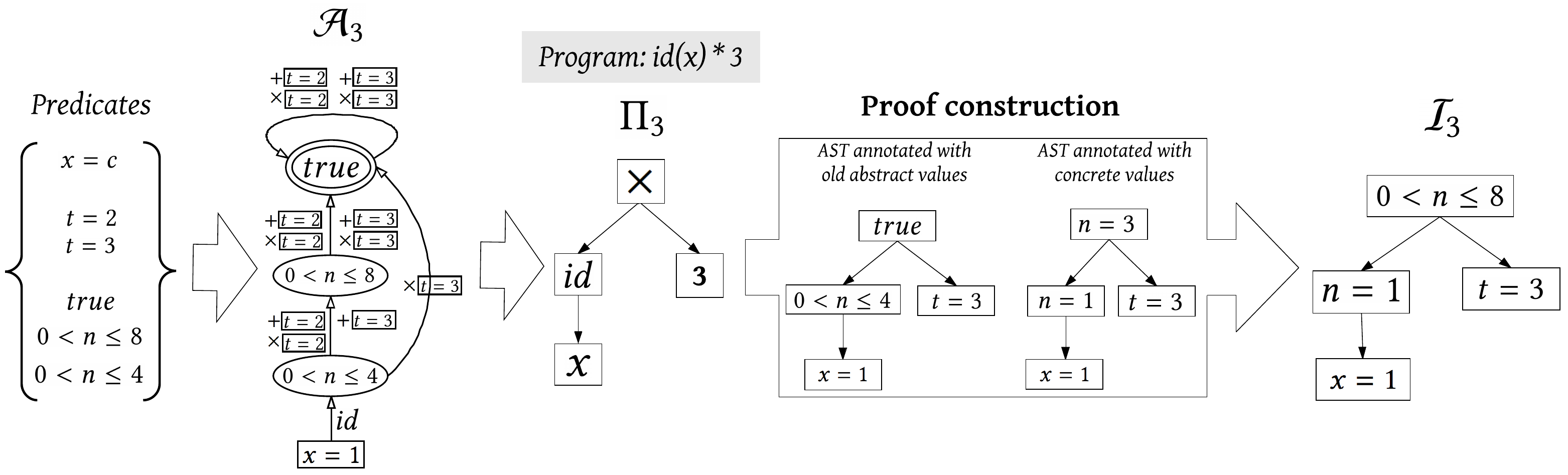}
\vspace{-5pt}
\caption*{Iteration 3: The constructed AFTA is $\fta_3$, {\sc Rank} returns $\prog_3$, $\prog_3$ is spurious, and the proof of incorrectness is $\itp_3$.}
\vspace{10pt}
\end{minipage}
\begin{minipage}{\linewidth}
\includegraphics[scale=0.43]{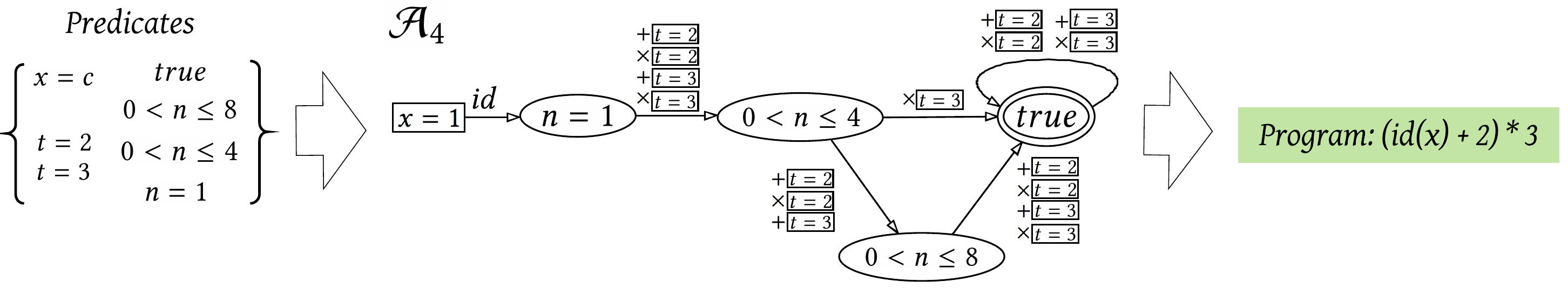}
\vspace{-5pt}
\caption*{Iteration 4: The constructed AFTA is $\fta_4$, and {\sc Rank} returns the desired program.}
\end{minipage}
\caption{Illustration of the synthesis algorithm.}
\figlabel{workingexample}
\vspace{-10pt}
\end{figure*}

\figref{workingexample} illustrates all iterations of the synthesis algorithm until we find the desired program. Let us now consider \figref{workingexample} in more detail.

\vspace{0.05in} \noindent
{\bf \underline{\emph{Iteration 1.}}}
As explained in \exref{afta}, the initial AFTA $\fta_1$ constructed by our algorithm accepts all DSL programs that start with $\emph{id}(x)$. Hence, in the first iteration, we obtain the program $\prog_1 = id(x)$ as a candidate solution. Since this program does not satisfy the example $1 \mapsto 9$, we construct a proof of incorrectness $\itp_1$, which introduces a new abstract value $0 < n \leq 8$ in our set of predicates.

\vspace{0.05in} \noindent
{\bf \underline{\emph{Iteration 2.}}} During the second iteration, we construct the AFTA labeled as $\fta_2$ in \figref{workingexample}, which contains a new state $0 < n \leq 8$. While $\fta_2$ no longer accepts the program $\emph{id}(x)$, it does accept the spurious program $\prog_2 = \emph{id}(x)+2$, which is returned by the ranking algorithm. Then we construct the proof of incorrectness for $\prog_2$, and we obtain a new predicate $0 < n \leq 4$.


\vspace{0.05in} \noindent
{\bf \underline{\emph{Iteration 3.}}} In the next iteration, we construct the AFTA labeled as $\fta_3$. Observe that $\fta_3$ no longer accepts the spurious program $\prog_2$ and also rules  out two other programs, namely  $id(x) + 3$ and $id(x) \times 2$. {\sc Rank} now returns the program  $\prog_3 = id(x) \times 3$, which is again spurious. After constructing the proof of incorrectness of $\prog_3$, we now obtain a new predicate $n=1$.


\vspace{0.05in} \noindent
{\bf \underline{\emph{Iteration 4.}}} In the final iteration, we construct the AFTA labeled as $\fta_4$, which rules out all programs containg a single operator ($+$ or $\times$) as well as 12 programs that use two operators. 
When we run the ranking algorithm on $\fta_4$, we obtain the candidate program $(id(x) + 2) \times 3$, which is indeed consistent with the example $1 \mapsto 9$. Thus,  the synthesis algorithm terminates with the solution  $(id(x) + 2) \times 3$.

\vspace{0.05in} \noindent
{\bf {\emph{Discussion.}}} As this example illustrates, our approach   explores far fewer programs compared to enumeration-based techniques. For instance, our algorithm only tested \emph{four} candidate programs against the input-output examples, whereas an enumeration-based approach would need to explore 24 programs. However, since each candidate program is generated using abstract finite tree automata, each iteration has a higher overhead. In contrast, the CFTA-based approach discussed in Section~\ref{sec:cfta} \emph{always} explores a single program, but the corresponding finite tree automaton may be \emph{very} large. Thus, our technique can be seen as providing a useful tuning knob between enumeration-based synthesis algorithms and representation-based techniques (e.g., CFTAs and \emph{version space algebras}) that construct a data structure representing all programs consistent with the input-output examples.

%% file: impl.tex
\section{Implementation and Instantiations}

We have implemented the synthesis algorithm proposed in this paper in a framework called \tool, written in Java.
\tool is parametrized over a DSL and its abstract semantics.
We have also instantiated \tool for two different domains, string transformation and matrix reshaping. In what follows, we describe our implementation of the \tool framework and its instantiations.
 
\subsection{Implementation of \tool Framework}\label{sec:implrank}

Our implementation of the \tool framework consists of three main modules, namely FTA construction, ranking algorithm, and proof generation. Since our implementation of FTA construction and proof generation mostly follows our technical presentation, we only focus on the implementation of the ranking algorithm, which is used to find a ``best" program that is accepted by the FTA. Our heuristic ranking algorithm returns a \emph{minimum-cost} AST accepted by the FTA, where the cost of  an AST  is defined as follows:
\[ 
\small 
\begin{array}{lll}
Cost(\emph{Leaf}(t)) & = & Cost(t) \\
Cost(\emph{Node}(f, \vec{\Pi})) & = & Cost(f) + \sum_{i} Cost(\Pi_i)
\end{array}
\]

In the above definition, $\emph{Leaf}(t)$ represents a leaf node of the AST labeled with terminal $t$, and $\emph{Node}(f, \vec{\Pi})$ represents a non-leaf node labeled with DSL  operator $f$ and subtrees $\vec{\Pi}$. Observe that the cost of an AST is calculated using the costs of DSL operators and terminals, which can be provided by the domain expert.
 
In our implementation, we identify a minimum-cost AST accepted by a finite tree automaton using the algorithm presented by~\citet{hypergraph} for finding a \emph{minimum weight} B-path in a weighted hypergraph. In the context of the ranking algorithm, we view an FTA as a hypergraph where states correspond to nodes and a transition $f(q_1, \ldots, q_n) \rightarrow q$ represents a B-arc $( \{ q_1, \dots, q_n \}, \{ q \} )$ where the weight of the arc is given by the cost of DSL operator $f$. We also add a dummy node $r$ to the hypergraph and an edge with weight $\emph{cost}(s)$ from $r$ to every node labeled $q_s^c$ where $s$ is a terminal symbol in the grammar. Given such a hypergraph representation of the FTA, the minimum-cost AST accepted by the FTA corresponds to a minimum-weight B-path from the dummy node $r$ to a node representing a final state in the FTA. 

\subsection{Instantiating \tool for String Transformations}\label{sec:inst-string}
 
To instantiate the \tool framework for a specific domain, the domain expert needs to provide a (cost-annotated) domain-specific language, a universe of possible predicates to be used in the abstraction, the abstract semantics of each DSL construct, and optionally an initial abstraction to use when constructing the initial AFTA. We now describe our instantiation of the \tool framework for synthesizing string transformation programs. 
 
\vspace{0.1in}  
{\noindent \bf \emph{Domain-specific language.}} 
Since there is significant prior work on automating string transformations using PBE~\cite{flashfill,blinkfill,flashmeta},  we directly adopt the DSL presented by \citet{blinkfill} as shown in \figref{stringdsl}.  This DSL essentially allows concatenating substrings of the input string $x$, where each substring is extracted using a start position $p_1$ and an end position $p_2$. A position can either be a constant index ($\texttt{ConstPos}(k)$) or the (start or end) index of the $k$'th occurrence of the match of token $\tau$ in the input string ($\texttt{Pos}(x, \tau, k, d)$). 

\begin{figure}
\small
\[
\begin{array}{rrll}
\text{String \ expr} & e & \assign & \texttt{Str}(f) \ | \ \texttt{Concat}(f, e); \\
\text{Substring \ expr} & f & \assign & \texttt{ConstStr}(s) \ | \ \texttt{SubStr}(x, p_1, p_2); \\
\text{Position} & p & \assign & \texttt{Pos}(x, \tau, k, d) \ | \ \texttt{ConstPos}(k); \\
\text{Direction} & d & \assign & \texttt{Start} \ | \ \texttt{End} ; \\
\end{array}
\]
\vspace{-10pt}
\caption{DSL for string transformations where $\tau$ represents a token, $k$ is an integer, and $s$ is a string constant.}
\figlabel{stringdsl}
\vspace{-10pt}
\end{figure}

\begin{figure}
\footnotesize 
\[
\begin{array}{rll}
\asemantics{f( s_1= c_1, \ldots, s_n=c_n )} & \assign & \big( s = \semantics{f(c_1, \ldots, c_n)} \big) 
\\ 
\asemantics{\texttt{Concat}( \emph{len}(f) = i_1, \emph{len}(e) = i_2 )} & \assign & \big( \emph{len}(e) = ( i_1 + i_2 ) \big) \\
\asemantics{\texttt{Concat}( \emph{len}(f) = i_1, e[i_2] = c )} & \assign & \big( e[i_1 + i_2] = c \big) \\
\asemantics{\texttt{Concat}( \emph{len}(f) = i, e = c )} & \assign & \big( \emph{len}(e) = (i + \emph{len}(c)) \wedge \bigwedge_{j = 0, \dots, \emph{len}(c) - 1}{e[ i + j ] = c[j]} \big) \\
\asemantics{\texttt{Concat}( f[i] = c, \pred) } & \assign & \big( e[i] = c \big) \\
\asemantics{\texttt{Concat}( f = c, \emph{len}(e) = i )} & \assign & \big( \emph{len}(e) = (\emph{len}(c) + i) \wedge \bigwedge_{j = 0, \dots, \emph{len}(c) - 1}{e[j] = c[j]} \big) \\
\asemantics{\texttt{Concat}( f = c_1, e[i] = c_2 )} & \assign & \big( e[\emph{len}(c_1) + i] = c_2 \wedge \bigwedge_{j = 0, \dots, \emph{len}(c_1) - 1}{e[j] = c_1[j]} \big) \\
\asemantics{\texttt{Str}( \pred )} & \assign & \pred \\ \\
\end{array}
\]
\vspace{-20pt}
\caption{Abstract semantics for the DSL shown in \figref{stringdsl}.}
\figlabel{stringdslasemantics}
\vspace{-10pt}
\end{figure}

\vspace{0.05in}
{\noindent \bf \emph{Universe.}}  
A natural abstraction when reasoning about strings is to consider their length; hence, our universe of predicates in this domain includes predicates of the form $\emph{len}(s) = i$, where $s$ is a symbol of type string and $i$ represents any integer. We also consider predicates of the form $s[i] = c$ indicating that the $i$'th character in string $s$ is $c$. Finally, recall from Section~\ref{sec:afta} that our universe must include predicates of the form $s = c$, where $c$ is a concrete value that symbol $s$ can take. Hence, our universe of predicates for the string domain is given by:
\[
\small 
\universe = \big \{ \emph{len}(s) =i  \ | \ i \in \mathbb{N} \big \} \cup  \big\{  s[ i] = c   \ | \ i \in \mathbb{N}, c \in \texttt{Char} ] \big\} \cup \big\{ s = c \ | \ c \in \texttt{Type}(s) \big\} \cup \ \big\{ \emph{true}, \emph{false} \big\}
\]

{\noindent \bf \emph{Abstract semantics.}} Recall from Section~\ref{sec:afta} that the DSL designer must provide an abstract transformer $\asemantics{f(\aval_1, \ldots, \aval_n)}$ for each grammar production $s \rightarrow f(s_1, \ldots, s_n)$  and abstract values $\aval_1, \ldots, \aval_n$. Since our universe of predicates can be viewed as the union of three different abstract domains for reasoning string length, character position, and string equality, our abstract transformers effectively define the reduced product of these abstract domains. In particular, we define a generic transformer for conjunctions of predicates as follows:
\[
\small 
f \big( (\wedge_{i_1} {\pred_{i_1}} ), \dots, (\wedge_{i_n} {\pred_{i_n}}) \big) \assign \sqcap_{i_1} \dots \sqcap_{i_n} {f(\pred_{i_1}, \dots, \pred_{i_n})}
\]

Hence, instead of defining a transformer for every possible abstract value (which may have arbitrarily many conjuncts),  it suffices to define an abstract transformer for every combination of atomic predicates.  We show the abstract transformers for all possible combinations of atomic predicates in \figref{stringdslasemantics}.

\vspace{0.05in}
{\noindent \bf \emph{Initial abstraction.}}  
Our initial abstraction includes predicates of the form $\emph{len}(s) = i$, where $s$ is a symbol of type string and $i$ is an integer, as well as the predicates in the default initial abstraction (see Section~\ref{sec:topsynthesis} for the definition).

\subsection{Instantiating \tool for Matrix and Tensor Transformations}\label{sec:inst-matrix}

Motivated by the abundance of  questions on how to perform various matrix and tensor transformations in MATLAB, we also use the \tool framework to synthesize tensor manipulation programs.\footnote{Tensors are generalization of matrices from $2$ dimensions to an arbitrary number of dimensions.} We believe this application domain is a good stress test for the \tool framework because 
(a) tensors are complex data structures which makes the search space larger, and 
(b) the input-output examples in this domain are typically much larger in size. Finally, we wish to show that the \tool framework can be immediately used to generate a practical synthesis tool for a new unexplored domain by providing a suitable DSL and its abstract semantics.

\begin{figure}[!t]
\[
\small 
\begin{array}{rrll}
\text{Tensor \ expr} & t & \assign & \texttt{id}(x) \ | \ \texttt{Reshape}(t, v) \ | \ \texttt{Permute}(t, v) \ | \ \texttt{Fliplr}(t) \ | \ \texttt{Flipud}(t); \\ 
\text{Vector \ expr} & v & \assign & [k_1, k_2] \ | \ \texttt{Cons}(k, v); \\
\end{array}
\]
\vspace{-10pt}
\caption{DSL for matrix transformations where $k$ is an integer.}
\figlabel{matrixdsl}
\vspace{-10pt}
\end{figure}

\begin{figure}
\footnotesize 
\[
\begin{array}{rll}
\asemantics{f( s_1= c_1, \ldots, s_n=c_n )} & \assign & \big( s = \semantics{f(c_1, \ldots, c_n)} \big) \\
\asemantics{\texttt{Cons}( k = i_1, \emph{len}(v) = i_2 )} & \assign & \big( \emph{len}(v) = (i_2 + 1)  \big) \\
\asemantics{\texttt{Permute}( \emph{numDims}(t) = i, p )} & \assign & \big( \emph{numDims}(t) = i \big) \\
\asemantics{\texttt{Permute}( \emph{numElems}(t) = i, p )} & \assign & \big( \emph{numElems}(t) = i \big) \\
\asemantics{\texttt{Reshape}( \emph{numDims}(t) = i_1, \emph{len}(v) = i_2 )} & \assign & \big( \emph{numDims}(t) = i_2 \big) \\
\asemantics{\texttt{Reshape}( \emph{numDims}(t) = i, v = c )} & \assign & \big( \emph{numDims}(t) = \emph{len}(c) \big) \\
\asemantics{\texttt{Reshape}( \emph{numElems}(t) = i, p )} & \assign & \big( \emph{numElems}(t) = i \big) \\
\asemantics{\texttt{Reshape}( t = c, \emph{len}(v) = i )} & \assign & \big( \emph{numElems}(t) = \emph{numElems}(c) \big)  \\
\asemantics{\texttt{Flipud}(p)} & \assign & p \\
\asemantics{\texttt{Fliplr}(p)} & \assign & p \\
\\ 
\end{array}
\]
\vspace{-20pt}
\caption{Abstract semantics the DSL shown in \figref{matrixdsl}.}
\figlabel{matrixdslasemantics}
\end{figure}

\vspace{0.05in}
{\noindent \bf \emph{Domain-specific language.}}  Our DSL for the tensor domain is inspired by existing MATLAB functions and is shown in \figref{matrixdsl}. 
In this DSL, tensor operators include \texttt{Reshape}, \texttt{Permute}, \texttt{Fliplr}, and \texttt{Flipud} and correspond to their namesakes in MATLAB\footnote{See the MATLAB documentation https://www.mathworks.com/help/matlab/ref/x.html where x refers to the name of the corresponding function.}.   For example, \texttt{Reshape}($t$, $v$) takes a tensor $t$ and a size vector $v$ and reshapes $t$ so that its dimension becomes $v$. Similarly, \texttt{Permute}($t, v$) rearranges the dimensions of tensor $t$ so that they are in the order specified by vector $v$. Next, \texttt{fliplr($t$)} returns tensor $t$ with its columns flipped in the left-right direction, and \texttt{flipud}($t$) returns tensor $t$ with its rows flipped in the up-down direction. Vector expressions are constructed recursively using the \texttt{Cons}($k,v$) construct, which yields a vector with first element $k$ (an integer), followed by  elements in vector $v$. 

\begin{example}
Suppose that we have a vector $v$ and we would like to reshape it in a row-wise manner so that it yields a matrix with $2$ rows and $3$ columns\footnote{StackOverflow post link: 
https://stackoverflow.com/questions/16592386/reshape-matlab-vector-in-row-wise-manner.}. For example, if the input vector is $[1,2,3,4,5,6]$, then we should obtain the matrix $[1,2,3; 4,5,6]$ where the semi-colon indicates a new row. This transformation can be expressed by the DSL program $\texttt{Permute}(\texttt{Reshape}( v, [3,2] ), [2,1] )$.
\end{example}

{\noindent \bf \emph{Universe of predicates.}}  Similar to the string domain, a  natural abstraction for vectors is to consider their length. Therefore, our universe includes predicates of the form $\emph{len}(v) = i$, indicating that vector $v$ has length $i$. In the case of tensors, our abstraction keeps track of the number of elements and number of dimensions of the tensors. In particular, the predicate $\emph{numDims}(t) = i$ indicates that $t$ is an $i$-dimensional tensor. Similarly, the predicate $\emph{numElems}(t) = i$ indicates that tensor $t$ contains a total of $i$ entries. Thus, the universe of predicates is given by:
\[ 
\universe = 
\left . 
\begin{array}{c}
\big\{ \emph{numDims}(t) = i \ | \ i \in \mathbb{N} \big\} \cup \big\{ \emph{numElems}(t) = i \ | \ i \in \mathbb{N} \big\}  \\
\big\{ \emph{len}(v) = i \ | \  i \in \mathbb{N} \big\}  \cup  \big\{ s = c \ | \ c \in \texttt{Type}(s) \big\} \ \cup \ \big\{ \emph{true}, \emph{false} \big\}
\end{array}
\right .
\]

\vspace{0.05in}
{\noindent \bf \emph{Abstract semantics.}}  The abstract transformers for all possible  combinations of atomic predicates for the DSL constructs are given in \figref{matrixdslasemantics}. As in the string domain, we  define a generic transformer for conjunctions of predicates as follows:
\[ 
\small 
f \big( ( \wedge_{i_1} {\pred_{i_1}} ), \dots, ( \wedge_{i_n} {\pred_{i_n}} ) \big) \assign \sqcap_{i_1} \dots \sqcap_{i_n} {f(\pred_{i_1}, \dots, \pred_{i_n})}
\]

\vspace{0.05in}
{\noindent \bf \emph{Initial abstraction.}} 
We do not specify the initial abstraction for matrix domain. That is, we use the default initial abstraction (see Section~\ref{sec:topsynthesis} for the definition). 

%% file: eval.tex
\section{Evaluation}\label{sec:eval}

We evaluate \tool by using it to automate string and matrix manipulation tasks collected from on-line forums and existing PBE benchmarks. 
The goal of our evaluation is to answer the following questions:
\begin{itemize}
\item {\bf Q1:} How does \tool perform on different synthesis tasks from the string and matrix domains?
\item {\bf Q2:} How many refinement steps does \tool take to find the correct program?
\item {\bf Q3:} What percentage of its running time does \tool spend in FTA vs. proof construction?
\vspace{2pt}
\item {\bf Q4:} How does \tool compare with existing synthesis techniques?
\item {\bf Q5:} What is the benefit of performing abstraction refinement in practice?
\end{itemize}


\subsection{Results for the String Domain}
In our first experiment, we evaluate \tool on \emph{all} 108 string manipulation benchmarks from the PBE track of the SyGuS competition~\cite{sygus}. We believe that the string domain is a good testbed for evaluating \tool because of the existence of mature  tools like FlashFill~\cite{flashfill} and the presence of a SyGuS benchmark suite for string transformations.

\vspace{0.03in}
\noindent 
{\bf \emph{Benchmark information.}}
Among the 108 SyGuS benchmarks related to string transformations, the number of examples range from 4 to 400, with an average of 78.2 and a median of 14. 
The average input example string length is 13.6 and the median is 13.0. The maximum (resp. minimum) string length is 54 (resp. 8). 

\vspace{0.03in}
\noindent 
{\bf \emph{Experimental setup.}}
We evaluate \tool using the string manipulation DSL shown in \figref{stringdsl} and the predicates and abstract semantics from Section \ref{sec:inst-string}. For each benchmark, we provide \tool with all 
input-output examples at the same time.\footnote{However, \tool typically uses a fraction of these examples when performing abstraction refinement.} 
We also compare \tool with the following existing synthesis techniques: 

\begin{itemize}
\item 
{\bf FlashFill:} This tool is the state-of-the-art synthesizer for automating string manipulation tasks and is shipped in Microsoft PowerShell as the ``convert-string'' commandlet. It propagates examples backwards using the inverse semantics of DSL operators, and adopts the VSA data structure to compactly represent the search space. 
\item 
{\bf \enum:} This technique based on enumerative search has been adopted to solve different kinds of synthesis problems~\cite{escher, transit, cheung2012, sygus}. It enumerates programs according to their size, groups them into equivalence classes based on their (concrete) input-output behavior to compress the search space, and returns the first program that is consistent with the examples. 
\item 
{\bf CFTA:} This is an implementation of the synthesis algorithm presented in Section \ref{sec:background}. It uses the concrete semantics of the DSL operators to construct an FTA whose language is exactly the set of programs that are consistent with the input-output examples.
\end{itemize}

To allow a fair comparison, we evaluate \enum and CFTA using  the same DSL and ranking heuristics that we use to evaluate \tool. For FlashFill, we use the ``convert-string'' commandlet from Microsoft Powershell that uses the same DSL.

Because the baseline techniques mentioned above perform \emph{much better} when the examples are provided in an interactive fashion~\footnote{Because \tool is not very sensitive to the number of examples, we used \tool in a non-interactive mode by providing all examples at once. Since the baseline tools do not scale as well in the number of examples, we used them in an interactive mode, with the goal of casting them in the best light possible.}, we evaluate them in the following way: Given a set of examples $E$ for each benchmark, we first sample an example $e$ in $E$, use each technique to synthesize a program $P$ that satisfies $e$, and check if $P$ satisfies all examples in $E$. If not, we sample another example $e'$ in $E$ for which $P$ does not produce the desired output, and repeat the synthesis process using both $e$ and $e^\prime$. The synthesizer terminates when it either successfully learns a program that meets all examples, proves that no program in the DSL satisfies the examples, or times out in 10 minutes.

\vspace{0.03in}
\noindent
{\bf \emph{\tool results.}}
\figref{blazestring} summarizes the results of our evaluation of \tool in the string domain. Because it is not feasible to give statistics for all 108 SyGuS benchmarks, we only show the detailed results for one benchmark from each of the 27 categories. Note that the four benchmarks within a category are very similar and only differ in the number of provided examples. The main take-away message from our evaluation is that \tool can successfully solve 70\% of the benchmarks in under a second, and 85\% of the benchmarks in under 4 seconds, with a median running time of 0.14 seconds. In comparison, the best solver, i.e., EUSolver~\cite{eusolver}, in the SyGuS'16 competition is able to solve in total 45 benchmarks within the timeout of 60 minutes~\cite{syguscomp16}. 

For most benchmarks, \tool spends the majority of its running time on FTA construction, whereas the time on proof construction is typically negligible. This is because the number of predicates that are considered in the proof construction phase is usually quite small. It takes \tool an average of 74 refinement steps before it finds the correct program. However, the median number of refinement steps is much smaller (17). Furthermore, as expected, there is a clear correlation between the number of iterations and total running time. Finally, we can observe that the synthesized programs are non-trival, with an average size of 25 in terms of the number of AST nodes.

\newcolumntype{?}{!{\vrule width .9pt}}
\newcommand{\anewline}{}

\begin{figure*}
\scriptsize  
\centering
\begin{tabular}{ c c ? c c c c c | c c c c c }
  \multicolumn{1}{c}{\emph{Benchmark}} 
& \multicolumn{1}{c ?}{$|\exs|$}  
& \multicolumn{1}{c}{$\emph{T}_{\emph{syn}}$ (sec)} 
& \multicolumn{1}{c}{$\emph{T}_{\fta}$} 
& \multicolumn{1}{c}{$\emph{T}_{\emph{rank}}$} 
& \multicolumn{1}{c}{$\emph{T}_{\itp}$} 
& \multicolumn{1}{c |}{$\emph{T}_{\emph{other}}$} 
& \multicolumn{1}{c}{\#\emph{Iters}} 
& \multicolumn{1}{c}{$|\ftastates_{\emph{final}}|$} 
& \multicolumn{1}{c}{$|\transitions_{\emph{final}}|$} 
& \multicolumn{1}{c}{$|\prog_{\emph{syn}}|$} \\ \hline\hline 
bikes & 6 & 0.05 & 0.05 & 0.00 & 0.00 & 0.00 & 1 & 52 & 135 & 13 \\ \anewline 
dr-name & 4 & 0.16 & 0.09 & 0.02 & 0.01 & 0.04 & 17 & 95 & 513 & 19 \\ \anewline 
firstname & 4 & 0.08 & 0.08 & 0.00 & 0.00 & 0.00 & 1 & 71 & 350 & 13 \\ \anewline 
initials & 4 & 0.11 & 0.09 & 0.00 & 0.01 & 0.01 & 14 & 68 & 209 & 32 \\ \anewline 
lastname & 4 & 0.10 & 0.10 & 0.00 & 0.00 & 0.00 & 3 & 79 & 450 & 13 \\ \anewline 
name-combine-2 & 4 & 0.20 & 0.12 & 0.02 & 0.01 & 0.05 & 45 & 101 & 549 & 32 \\ \anewline 
name-combine-3 & 6 & 0.16 & 0.10 & 0.01 & 0.02 & 0.03 & 26 & 80 & 305 & 32 \\ \anewline 
name-combine-4 & 5 & 0.30 & 0.14 & 0.03 & 0.05 & 0.08 & 62 & 114 & 725 & 35 \\ \anewline 
name-combine & 6 & 0.16 & 0.10 & 0.02 & 0.02 & 0.02 & 20 & 87 & 427 & 29 \\ \anewline 
phone-1 & 6 & 0.07 & 0.07 & 0.00 & 0.00 & 0.00 & 2 & 43 & 79 & 13 \\ \anewline 
phone-10 & 7 & 1.99 & 0.69 & 0.34 & 0.30 & 0.66 & 539 & 471 & 4754 & 48 \\ \anewline 
phone-2 & 6 & 0.06 & 0.06 & 0.00 & 0.00 & 0.00 & 3 & 43 & 77 & 13 \\ \anewline 
phone-3 & 7 & 0.25 & 0.12 & 0.03 & 0.05 & 0.05 & 59 & 88 & 355 & 35 \\ \anewline 
phone-4 & 6 & 0.23 & 0.10 & 0.03 & 0.04 & 0.06 & 63 & 155 & 1256 & 45 \\ \anewline 
phone-5 & 7 & 0.08 & 0.08 & 0.00 & 0.00 & 0.00 & 1 & 53 & 114 & 13 \\ \anewline 
phone-6 & 7 & 0.10 & 0.10 & 0.00 & 0.00 & 0.00 & 2 & 53 & 112 & 13 \\ \anewline 
phone-7 & 7 & 0.08 & 0.08 & 0.00 & 0.00 & 0.00 & 3 & 53 & 108 & 13 \\ \anewline 
phone-8 & 7 & 0.11 & 0.11 & 0.00 & 0.00 & 0.00 & 4 & 53 & 106 & 13 \\ \anewline 
phone-9 & 7 & 1.09 & 0.34 & 0.19 & 0.15 & 0.41 & 269 & 454 & 7355 & 61 \\ \anewline 
phone & 6 & 0.07 & 0.07 & 0.00 & 0.00 & 0.00 & 1 & 43 & 80 & 13 \\ \anewline 
reverse-name & 6 & 0.14 & 0.08 & 0.01 & 0.02 & 0.03 & 20 & 83 & 414 & 29 \\ \anewline 
univ\_1 & 6 & 1.34 & 0.61 & 0.21 & 0.12 & 0.40 & 149 & 348 & 9618 & 32 \\ \anewline 
univ\_2 & 6 & T/O & --- & --- & --- & --- & --- & --- & --- & --- \\ \anewline 
univ\_3 & 6 & 3.69 & 1.63 & 0.57 & 0.15 & 1.34 & 405 & 467 & 18960 & 22 \\ \anewline 
univ\_4 & 8 & T/O & --- & --- & --- & --- & --- & --- & --- & --- \\ \anewline 
univ\_5 & 8 & T/O & --- & --- & --- & --- & --- & --- & --- & --- \\ \anewline 
univ\_6 & 8 & T/O & --- & --- & --- & --- & --- & --- & --- & --- \\ \hline\hline 
Median & 6 & 0.14 & 0.10 & 0.01 & 0.01 & 0.02 & 17 & 80 & 355 & 22 \\ \anewline 
Average & 6.1 & 0.46 & 0.22 & 0.06 & 0.04 & 0.14 & 74.3 & 137.1 & 2045.7 & 25.3 \\ \hline
\end{tabular}
\vspace{-0.1in}
\caption{\tool results for the string domain, where $|\vec{e}|$ shows the number of examples and $\emph{T}_\emph{syn}$ gives  synthesis time in seconds. The next columns labeled $T_x$ show the time for FTA construction, ranking, proof construction, and all remaining parts (e.g. FTA minimization). \#\emph{Iters} shows the number of refinement steps, and  $|Q_\emph{final}|$ and $|\Delta_\emph{final}|$ show the number of states and transitions in the final AFTA. The last column labeled $|\prog_\emph{syn}|$ shows the size of the synthesized program (measured by number of AST nodes). The timeout is set to be 10 minutes.}
\vspace{-10pt}
\figlabel{blazestring}
\end{figure*}

\vspace{0.03in}
\noindent
{\bf \emph{Comparison.}}
\figref{stringcomparison} compares the running times of \tool with  FlashFill, \enum, and CFTA on \emph{all} 108 SyGuS benchmarks.
Overall, \tool solves the most number of benchmarks (90), with an average running time of 0.49 seconds. Furthermore, any benchmark that can be solved using FlashFill, \enum, or CFTA can also be solved by \tool.

Compared to CFTA, \tool solves 60\% more benchmarks (90 vs. 56) and outperforms CFTA by 363x (in terms of running time) on the 56  benchmarks that can be solved by both techniques.
This result demonstrates that  abstraction refinement  helps scale up  the CFTA-based synthesis technique to solve more benchmarks in much less time. 

Compared to \enum, the improvement of \tool is moderate for relatively simple benchmarks. In particular,  for the 40 benchmarks that \enum can solve in under 1 second, \tool (only) shows a 1.5x improvement in  running time. However, for more complex synthesis tasks, the performance of \tool is significantly better than \enum. For the 54 benchmarks that can be solved by both techniques, we observe a 16x improvement in  running time. Furthermore, \tool can solve 36 benchmarks on which \enum times out. We believe this result demonstrates the advantage of using abstract values for search space reduction. 


\begin{figure*}
\centering
\begin{minipage}{.75\linewidth}
\includegraphics[scale=0.46]{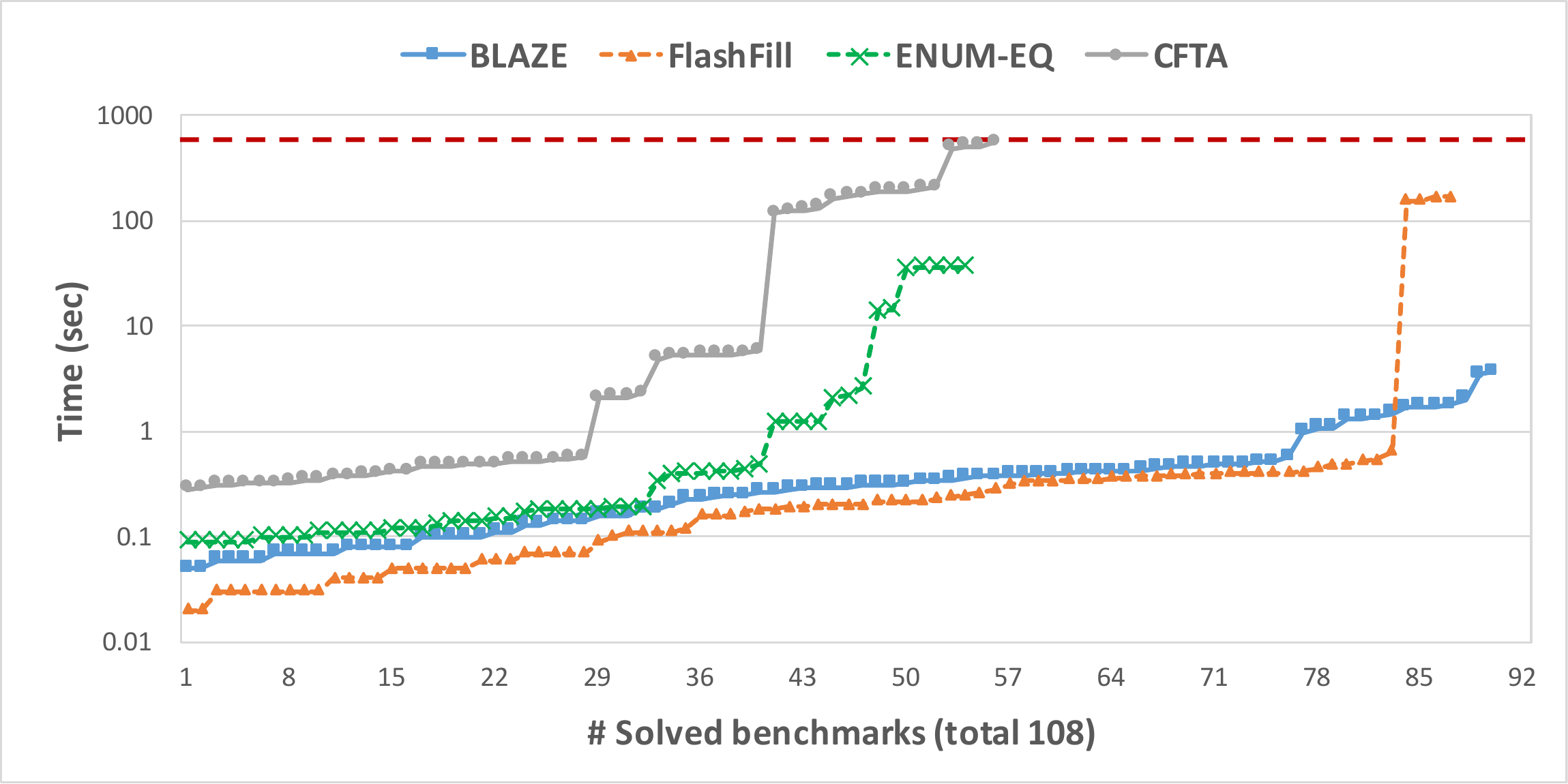}
\end{minipage}
\begin{minipage}{0.24\linewidth}
\footnotesize
\begin{tabular}{c | c c c }
\hline 
 & \rot{\# Solved} & \rot{ \ Average time (sec) \ }  \\ \hline 
\rule{0pt}{10pt} \tool & 90 & 0.49 \\ [2pt]
FlashFill & 87 & 7.66  \\ [2pt]
\enum & 54 & 4.25 \\ [2pt]
CFTA & 56 & 73.91  \\ [2pt] \hline 
\end{tabular}
\end{minipage}
\caption{Comparison with existing techniques. A data point $(X, Y)$ means that $X$ benchmarks are solved within a maximum running time of $Y$ seconds (per benchmark). The timeout is set to be 10 minutes.}
\figlabel{stringcomparison}
\vspace{-10pt}
\end{figure*}

Finally, \tool also compares favorably with FlashFill, the state-of-the-art technique for automating string transformation tasks. In particular, \tool achieves very competitive performance for the benchmarks that both techniques can solve. Furthermore, \tool can solve 3 benchmarks on which FlashFill times out. 
Since FlashFill is a domain-specific synthesizer that has been crafted specifically for automating string manipulation tasks, we believe these results demonstrate that \tool can compete with domain-specific state-of-the-art synthesizers.

\vspace{0.03in}
\noindent
{\bf \emph{Outlier analysis.}} All techniques, including \tool, time out on 18 benchmarks for the univ\_x category. We investigated the cause of failure for these benchmarks and found that the desired program for most of these benchmarks cannot be expressed in the underlying DSL.


\subsection{Results for the Matrix Domain}

In our second experiment, we evaluate \tool on matrix and tensor transformation benchmarks obtained from on-line forums. Because tensors are more complicated data structures than strings,  the search space in this domain tends to be larger on average compared to the string domain. Furthermore,  since automating matrix transformations is a  useful (yet unexplored) application of programming-by-example, we believe  this domain is an interesting target for \tool.

To perform our evaluation, we collected 39  benchmarks from two on-line forums, namely StackOverflow and MathWorks.\footnote{MathWorks (https://www.mathworks.com/matlabcentral/answers/) is a help forum for MATLAB users.} Our benchmarks were collected using the following methodology: We searched for the keyword \emph{"matlab matrix reshape''}  and then sorted the results according to their relevance. We then looked at the first 100 posts from each forum and retained  posts that contain at least one example as well as the target program is in one of the responses. 

\begin{figure*}
\scriptsize  
\centering
\begin{tabular}{ c ? c  c c  c  c | c c c c c }
  \multicolumn{1}{c ?}{\emph{Benchmark}} 
& \multicolumn{1}{c}{$\emph{T}_{\emph{syn}}$ (sec)} 
& \multicolumn{1}{c}{$\emph{T}_{\fta}$} 
& \multicolumn{1}{c}{$\emph{T}_{\emph{rank}}$} 
& \multicolumn{1}{c}{$\emph{T}_{\itp}$} 
& \multicolumn{1}{c |}{$\emph{T}_{\emph{other}}$} 
& \multicolumn{1}{c}{\#\emph{Iters}} 
& \multicolumn{1}{c}{$|\ftastates_{\emph{final}}|$} 
& \multicolumn{1}{c}{$|\transitions_{\emph{final}}|$} 
& \multicolumn{1}{c}{$|\prog_{\emph{syn}}|$} \\ \hline\hline 
stackoverflow-1 & 0.29 & 0.14 & 0.02 & 0.08 & 0.05 & 39 & 125 & 993 & 10 \\ \anewline 
stackoverflow-2 & 2.74 & 0.86 & 0.10 & 1.52 & 0.26 & 319 & 279 & 4483 & 22 \\ \anewline 
stackoverflow-3 & 0.72 & 0.20 & 0.03 & 0.43 & 0.06 & 57 & 143 & 1334 & 14 \\ \anewline 
stackoverflow-4 & 13.32 & 0.31 & 0.04 & 12.89 & 0.08 & 166 & 165 & 959 & 22 \\ \anewline 
stackoverflow-5 & 1.34 & 0.57 & 0.08 & 0.48 & 0.21 & 222 & 236 & 2595 & 18 \\ \anewline 
stackoverflow-6 & 0.42 & 0.17 & 0.02 & 0.17 & 0.06 & 48 & 129 & 1012 & 10 \\ \anewline 
stackoverflow-7 & 2.04 & 0.59 & 0.07 & 1.20 & 0.18 & 217 & 244 & 2607 & 18 \\ \anewline 
stackoverflow-8 & 2.04 & 0.83 & 0.08 & 0.90 & 0.23 & 288 & 280 & 3447 & 18 \\ \anewline 
stackoverflow-9 & 1.67 & 0.90 & 0.08 & 0.44 & 0.25 & 114 & 374 & 5389 & 16 \\ \anewline 
stackoverflow-10 & 0.23 & 0.12 & 0.01 & 0.06 & 0.04 & 28 & 114 & 715 & 10 \\ \anewline 
stackoverflow-11 & 0.74 & 0.34 & 0.05 & 0.24 & 0.11 & 106 & 155 & 1004 & 18 \\ \anewline 
stackoverflow-12 & 0.82 & 0.12 & 0.02 & 0.63 & 0.05 & 38 & 124 & 929 & 10 \\ \anewline 
stackoverflow-13 & 0.59 & 0.17 & 0.02 & 0.34 & 0.06 & 49 & 143 & 1227 & 12 \\ \anewline 
stackoverflow-14 & 52.94 & 1.36 & 0.11 & 51.24 & 0.23 & 385 & 324 & 4321 & 22 \\ \anewline 
stackoverflow-15 & 0.41 & 0.12 & 0.01 & 0.24 & 0.04 & 31 & 121 & 611 & 14 \\ \anewline 
stackoverflow-16 & 5.02 & 0.38 & 0.06 & 4.45 & 0.13 & 228 & 172 & 1083 & 22 \\ \anewline 
stackoverflow-17 & 2.54 & 0.79 & 0.09 & 1.42 & 0.24 & 319 & 279 & 4483 & 22 \\ \anewline 
stackoverflow-18 & 0.54 & 0.25 & 0.03 & 0.18 & 0.08 & 65 & 144 & 1201 & 14 \\ \anewline 
stackoverflow-19 & 0.73 & 0.36 & 0.06 & 0.17 & 0.14 & 142 & 162 & 1180 & 18 \\ \anewline 
stackoverflow-20 & 1.31 & 0.36 & 0.05 & 0.78 & 0.12 & 165 & 160 & 786 & 18 \\ \anewline 
stackoverflow-21 & 1.01 & 0.52 & 0.06 & 0.27 & 0.16 & 180 & 195 & 1566 & 18 \\ \anewline 
stackoverflow-22 & 0.21 & 0.10 & 0.01 & 0.07 & 0.03 & 19 & 106 & 526 & 10 \\ \anewline 
stackoverflow-23 & 1.24 & 0.26 & 0.04 & 0.85 & 0.09 & 108 & 181 & 2493 & 14 \\ \anewline 
stackoverflow-24 & 0.62 & 0.14 & 0.02 & 0.41 & 0.05 & 52 & 138 & 1183 & 12 \\ \anewline 
stackoverflow-25 & 0.81 & 0.20 & 0.03 & 0.51 & 0.07 & 72 & 170 & 2201 & 14 \\ \anewline 
mathworks-1 & 0.71 & 0.15 & 0.02 & 0.48 & 0.06 & 55 & 137 & 1103 & 12 \\ \anewline 
mathworks-2 & 0.88 & 0.11 & 0.02 & 0.71 & 0.04 & 34 & 126 & 848 & 14 \\ \anewline 
mathworks-3 & 1.07 & 0.58 & 0.06 & 0.27 & 0.16 & 180 & 195 & 1566 & 18 \\ \anewline 
mathworks-4 & 3.94 & 0.22 & 0.03 & 3.62 & 0.07 & 89 & 195 & 2589 & 14 \\ \anewline 
mathworks-5 & 0.45 & 0.15 & 0.02 & 0.22 & 0.06 & 45 & 134 & 963 & 12 \\ \anewline 
mathworks-6 & 1.30 & 0.42 & 0.07 & 0.63 & 0.18 & 195 & 222 & 2100 & 18 \\ \anewline 
mathworks-7 & 0.21 & 0.10 & 0.01 & 0.06 & 0.04 & 28 & 116 & 717 & 10 \\ \anewline 
mathworks-8 & 0.27 & 0.13 & 0.02 & 0.07 & 0.05 & 39 & 125 & 993 & 10 \\ \anewline 
mathworks-9 & 1.73 & 0.23 & 0.03 & 1.39 & 0.08 & 104 & 160 & 955 & 10 \\ \anewline 
mathworks-10 & 1.57 & 0.30 & 0.05 & 1.10 & 0.12 & 145 & 172 & 1176 & 14 \\ \anewline 
mathworks-11 & 9.40 & 5.72 & 0.50 & 1.83 & 1.35 & 613 & 583 & 25924 & 22 \\ \anewline 
mathworks-12 & 1.25 & 0.36 & 0.07 & 0.66 & 0.16 & 187 & 203 & 1799 & 18 \\ \anewline 
mathworks-13 & 2.49 & 1.45 & 0.17 & 0.41 & 0.46 & 462 & 295 & 2574 & 15 \\ \anewline 
mathworks-14 & 11.10 & 6.18 & 1.19 & 0.60 & 3.13 & 827 & 678 & 34176 & 22 \\ \hline\hline 
Median & 1.07 & 0.30 & 0.04 & 0.48 & 0.09 & 108 & 165 & 1201 & 14 \\ \anewline 
Average & 3.35 & 0.67 & 0.09 & 2.36 & 0.23 & 165.6 & 205.2 & 3225.9 & 15.5 \\ \hline
\end{tabular}
\vspace{-0.1in}
\caption{\tool results for matrix domain. We use the same notation explained in the caption of \figref{blazestring}.}
\figlabel{blazematrix}
\vspace{-10pt}
\end{figure*}

\vspace{0.03in}
\noindent
{\bf \emph{Benchmark information.}}
Since the overwhelming majority of forum entries contain a single example, we only provide one input-output example for each benchmark. 
The number of entries in the input  tensor ranges from 6 to 640. The average number is 73.5, and the median is 36. Among all benchmarks, 29  involve transforming the input example into tensors of dimension great than 2.

\vspace{0.03in}
\noindent
{\bf \emph{Experimental setup.}}
We evaluate \tool using the DSL  shown in \figref{matrixdsl} and the abstract semantics presented in Section \ref{sec:inst-matrix}. 
Similar to the string domain, we also compare \tool with \enum and CFTA. However, since there is no existing domain-specific synthesizer for automating matrix transformation tasks, we implemented a specialized VSA-based synthesizer for our matrix domain by instantiating the \prose framework~\cite{flashmeta}. In particular, we provide precise witness functions for all the operators in our DSL, which allows \prose to effectively decompose the synthesis task. 
To allow a fair comparison, we use the same DSL for all the synthesizers, as well as the same ranking heuristics. 
We also experiment with all baseline synthesizers in the interactive setting, as we did for the string domain. 
The timeout is set to be 10 minutes.

\vspace{0.03in}
\noindent
{\bf \emph{\tool results.}}
The results of our evaluation on \tool are summarized in \figref{blazematrix}. As shown in the figure, \tool can successfully solve all benchmarks with an average (resp. median) synthesis time of 3.35 (resp. 1.07)  seconds. Furthermore, \tool can solve 46\% of the benchmarks in under 1 second, and 87\% of the benchmarks in under 5 seconds. These results demonstrate that \tool is also practical for automating matrix/tensor reshaping tasks.

Looking at \figref{blazematrix} in more detail,  \tool takes an average of 165 refinement steps to find a correct program. 
Unlike the string domain where \tool spends most of its time in FTA construction, proof construction also seems to take significant time in the matrix domain.  We conjecture this is because \tool needs to search for predicates in a large space. 
The final AFTA constructed by \tool contains an average of 205 states, and the average AST size of synthesized programs is 16.

\vspace{0.03in}
\noindent
{\bf \emph{Comparison.}}
As shown in \figref{matrixcomparison}, \tool significantly outperforms all existing techniques, both in terms of  the number of solved benchmarks as well as the running time. In particular, we observe a 262x improvement over CFTA, a 115x improvement over \enum, and a 90x improvement over \prose  in terms of the running time. Therefore, this experiment also demonstrates the advantage of  using abstract values and abstraction refinement in the matrix domain.

\begin{figure*}
\centering
\begin{minipage}{.73\linewidth}
\includegraphics[scale=0.48]{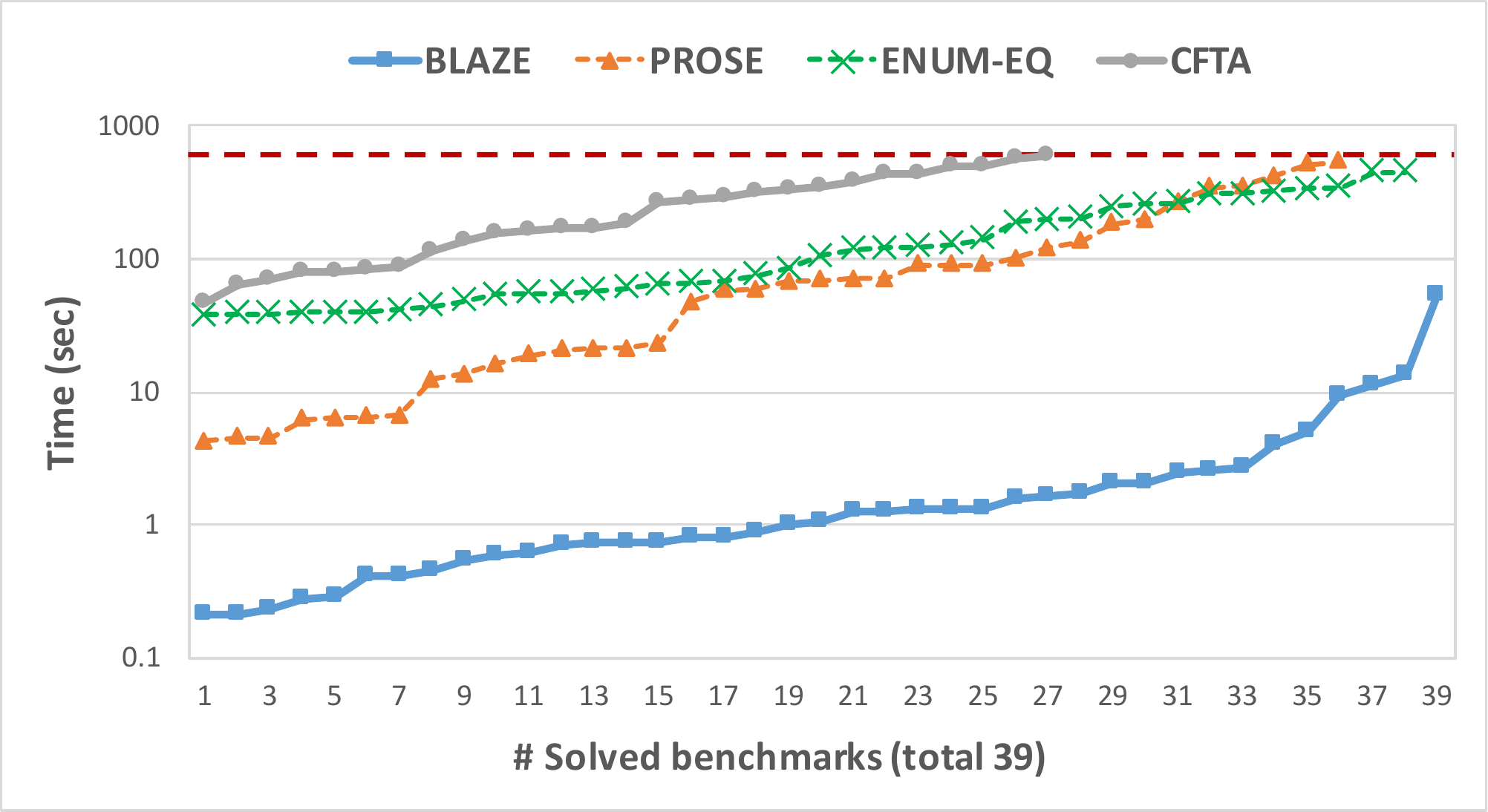}
\end{minipage}
\begin{minipage}{0.26\linewidth}
\footnotesize 
\begin{tabular}{c | c c c }
\hline 
 & \rot{\# Solved} & \rot{ \ Average time (sec) \ } \\ \hline 
\rule{0pt}{10pt} \tool & 39 & 3.35  \\ [2pt]
\prose & 36 & 113.13 \\ [2pt]
ENUM-EQ & 38 & 147.88 \\ [2pt]
CFTA & 27 & 252.80 \\ [2pt] \hline 
\end{tabular}
\end{minipage}
\vspace{-0.1in}
\caption{Comparison with existing techniques.}
\figlabel{matrixcomparison}
\vspace{-10pt}
\end{figure*}

\vspace{0.03in}
\noindent
{\bf \emph{Outlier analysis.}}
The benchmark named ``stackoverflow-14'' takes 53 seconds because the input example tensor is the largest one we have in our benchmark set (with 640 entries). As a result, in the proof construction phase \tool needs to search for the desired formula in a space that contains over $10^5$ conjunctions. This makes the synthesis process computationally expensive.

\subsection{Discussion}

The reader may wonder why \tool performs much better in the matrix domain compared to VSA-based techniques (FlashFill and Prose) than in the string domain. We conjecture that this discrepancy can be explained by considering the size of the search space measured in terms of the number of (intermediate) concrete values produced by the DSL programs. For the string domain, the search space size is dominated by the number of substrings, and FlashFill constructs $n^2$ nodes for substrings in the VSA data structure, where $n$ is the length of the output example.  For the matrix domain, the search space size is mostly determined by the number of intermediate matrices; in the worse case \prose would have to explore $\mathcal{O}(n!)$ nodes, where $n$ is the number of entries in the example matrix.
Hence, the size of the search space in the matrix domain is potentially much larger for VSA-based techniques than that in the string domain.  
In contrast, \tool performs quite well in both application domains, since it uses abstract values (instead of concrete values) to represent equivalence classes. 

%% file: related.tex
\section{Related work}\label{sec:related}
In this section, we compare our technique against related approaches in the synthesis and verification literature.

\vspace{0.06in}
{\noindent \bf \emph{CEGAR in model checking.}}  Our approach is  inspired by the use of counterexample-guided abstraction refinement in software model checking~\cite{blast1,blast2,blast3,slam}. The idea here is to start with a coarse  abstraction of the program and then perform model checking over this abstraction. Since any errors encountered using this approach may be spurious, the model checker then looks for a counterexample trace and refines the abstraction if the error is indeed spurious. While there are many ways to perform refinement, a popular approach is to refine the abstraction using \emph{interpolation}, which provides a proof of unsatisfiability of a trace~\cite{blast3}. Our approach is very similar  to CEGAR-based model checkers in that we perform abstraction refinement whenever we find a \emph{spurious program} as opposed to a spurious error trace. In addition, the proofs of incorrectness that we utilize in this paper can be viewed as a form of \emph{tree interpolant}~\cite{tree-interpolant1,tree-interpolant2}.

\vspace{0.06in}
{\noindent \bf \emph{Abstraction in Program Synthesis.}} The only prior work that uses abstraction refinement in the context of synthesis is the \emph{abstraction-guided synthesis} (AGS) technique of Vechev et al. for learning efficient synchronization in concurrent programs~\cite{ags}. Unlike \tool which aims to learn an entire program from scratch using input-output examples, AGS requires an input concurrent program and only performs small modifications to the program by adding synchronization primitives. In more detail, AGS first performs an abstraction of the program and checks whether there are any counterexample (abstract) interleavings that violate the given safety constraint. If there is no violation, it returns the current program. Otherwise, it non-deterministically chooses to either refine the abstraction or modify the program by adding synchronization primitives such that the abstract interleaving is removed.  AGS can be viewed as a program repair technique for concurrent programs and cannot be used for synthesizing programs from input-output examples.

Other synthesizers that bear similarities to the approach proposed in this paper include {\sc Synquid}~\cite{synquid} and {\sc Morpheus}~\cite{morpheus}. In particular, both of these techniques use specifications of DSL constructs in the form of refinement types and first-order formulas respectively, and use these specifications to refute  programs that do not satisfy the specification. Similarly, \tool uses abstract semantics of DSL constructs, which can be viewed as specifications. However, unlike {\sc Synquid}, the specifications in \tool and {\sc Morpheus} overapproximate the behavior of the DSL constructs. \tool further differs from both of these techniques in that it performs abstraction refinement and learns programs using finite tree automata.

There is a line of work that uses abstractions in the context of component-based program synthesis~\cite{gascon2017look, Tiwari2015}. These techniques annotate each component with a ``decoration'' that serves as an abstraction of the semantics of that component. The use of such abstractions simplifies the synthesis task by reducing a complex $\exists \forall$ problem to a simpler $\exists \exists $ constraint solving problem, albeit at the cost of the completeness. In contrast to these techniques, our method uses abstractions to construct a more compact abstract FTA and performs abstraction refinement to rule out spurious programs. 

The use of abstraction refinement has also been explored in the context of superoptimizing compilers~\cite{phothilimthana16asplas}. In particular,  Phothilimthana et al. use test cases to construct an (underapproximate) abstraction of program behavior and ``refine'' this abstraction by iteratively including more test cases. However, since this abstraction is heuristically applied to ``promising'' parts of the candidate space, this method may not be able to find the desired equivalent program. This technique differs significantly from our method in that they use an orthogonal definition of abstraction and perform abstraction refinement in a different heuristic-guided manner.

Another related technique is Storyboard Programming~\cite{storyboard} for learning data structure manipulation programs from examples by combining abstract interpretation and shape analysis. However, it differs from \tool in that the user needs to manually provide precise abstractions for input-output examples as well as abstract transformers for data structure operations. Furthermore, there is no automated refinement phase.

\vspace{0.06in}
{\noindent \bf \emph{Programming-by-Example (PBE).}} The problem of automatically learning programs that are consistent with a set of input-output examples  has been the subject of research for the last four decades~\cite{shaw1975}. Recent advances in algorithmic and logical reasoning techniques have led to the development of PBE systems in several  domains including regular expression based string transformations~\cite{flashfill,blinkfill},  data structure manipulations~\cite{lambda2,hades}, network policies~\cite{yuan2014netegg}, data filtering~\cite{fidex}, file  manipulations~\cite{strisynth}, interactive parser synthesis~\cite{parsersynthesis}, and synthesizing map-reduce  distributed programs~\cite{mapreduce}. It has also been studied from different perspectives, such as type-theoretic interpretation~\cite{scherer2015, myth, myth2}, version space learning~\cite{flashmeta, flashfill}, and deep learning~\cite{neurosymbolic, robustfill}.


Our method, SYNGAR, presents a new approach to example-guided program synthesis using abstraction refinement. Unlike most of the earlier PBE approaches that prune the search space using the concrete semantics of DSL operators~\cite{escher,transit}, SYNGAR, instead, uses their abstract semantics  and iteratively refines the abstraction until it finds a program that satisfies the input-output examples. Although we instantiate SYNGAR in two domains, namely string and matrix transformations, we believe the SYNGAR approach can be used to complement many previous PBE systems to make  synthesis  more efficient.

\vspace{0.06in}
{\noindent \bf \emph{Counterexample-guided Inductive Synthesis (CEGIS).}} Counterexample guided inductive synthesis (\textsc{CEGIS})~\cite{comsketching,solarthesis} is a popular algorithm used for solving  synthesis problems of the form $\exists P ~ \forall i: \phi(P,i)$ where the goal is to synthesize a program $P$ such that the specification $\phi$ holds for all inputs $i$. The main idea of the algorithm is to reduce the solving of the second-order formula to two first-order formulas: i) $\exists P : \phi(P,{I_1,\cdots,I_k}) $ (synthesis), and ii) $\exists i : \neg\phi(P,i) $ (verification). The first phase learns a program $P$ that is consistent with a finite set of inputs (${I_1,\cdots,I_k}$), whereas the second phase performs verification on the learnt candidate program $P$ to find a counterexample input $i$ that violates the specification. If such an input $i$ exists, the input is added to the set of current inputs and the synthesis phase is repeated. This iterative process continues until either the verification check succeeds (i.e., the learnt program $P$ satisfies the specification) or if the synthesis check fails (i.e., there is no program that satisfies the specification). 

{\sc CEGIS} bears similarities to SYNGAR in that both approaches are guided by counterexamples (i.e., incorrect programs). However, the two approaches are very different in that {\sc CEGIS} performs abstraction over the specification, whereas SYNGAR performs abstraction over the DSL constructs. In particular, the input-output examples used in the synthesis phase of {\sc CEGIS} \emph{under-approximate} the specification, whereas the abstract finite tree automata in SYNGAR \emph{over-approximate} the set of programs that are consistent with the specification. Since  SYNGAR  is intended for example-guided synthesis, we believe that it can be used to complement the synthesis phase in {\sc CEGIS}.


\vspace{0.06in}
{\noindent \bf \emph{Finite Tree Automata (FTA).}} Tree automata, which generalize finite word automata, date back to 1968 and were originally used for proving the existence of a decision procedure for weak monadic second-order logic~\cite{thatcher1968}. Since then, tree automata have been found applications in the analysis of XML documents~\cite{fta-xml1,fta-xml2}, software verification~\cite{fta-verif1,fta-verif2,fta-verif3,fta-verif4} and natural language processing~\cite{fta-nlp1,fta-nlp2}. Recent work by Kafle and Gallagher is particularly related in that they use counterexample-guided abstraction refinement to solve a system of constrained Horn Clauses and perform refinement using finite tree automata~\cite{fta-verif3}. In contrast to their approach, we use finite tree automata for synthesis rather than for refinement.

Finite tree automata have also found interesting applications in the context of program synthesis. For example, Parthasarathy uses finite tree automata as a theoretical basis for reactive synthesis~\cite{fta-reactive}. Specifically, given an $\omega$-specification of the reactive system, their technique constructs a tree automaton that accepts all programs that meet the specification. Recent work by Wang et al. also uses finite tree automata for synthesizing data completion scripts from input-output examples~\cite{dace}. In this work, we generalize the technique of Wang et al. by showing how it can be used to synthesize programs over any arbitrary DSL described in a context-free grammar. We also introduce the concept of abstract finite tree automata (AFTA) and describe a method for counterexample-guided synthesis using AFTAs.

%% file: conc.tex
\section{Conclusion}

We proposed a new synthesis methodology, called SYNGAR, for synthesizing programs that are consistent with a given set of input-output examples. The key idea  is to find a program that satisfies the examples according to the abstract semantics of the DSL constructs and then check the correctness of the learnt program with respect to the concrete semantics. Our approach returns the synthesized program if it satisfies the examples, and automatically refines the abstraction otherwise. Our method performs synthesis using abstract finite tree automata and refines the abstraction by constructing a proof of incorrectness of the learnt program for a given input-output example.

We have implemented the proposed methodology in a synthesis framework called \tool and instantiated it for two different domains, namely string and matrix transformations. Our evaluation shows that \tool is competitive with FlashFill in the string domain and that it outperforms Prose by 90x in the matrix domain. Our evaluation also shows the advantages of using  abstract semantics and performing abstraction refinement. 

%% file: limitations.tex
\section{Limitations and Future Work}
While our proposed SYNGAR framework can be realized using different synthesis algorithms, the FTA-based method described in this paper has three key limitations: First, our synthesis method does not support let bindings, thus, it cannot be used to synthesize programs over DSLs that allow variable introduction. Second, our method treats $\lambda$-abstractions as constants; hence, it may not perform very well for DSLs that encourage heavy use of higher-order combinators. Third, our method requires abstract values to be drawn from a decidable logic -- i.e., we assume that it is  possible to over-approximate values of intermediate DSL expressions using formulas from a decidable logic. In future work, we plan to develop new abstract synthesis algorithms that support DSLs with richer language features, including let bindings. We also plan to explore more efficient techniques for synthesizing programs over DSLs that make heavy use of higher-order combinators.

%% file: ack.tex
\begin{acks}
We would like to thank members in the UToPiA group for their insightful comments, as well as the anonymous reviewers for their constructive feedback. This work was supported in part by NSF Award \#1712060, NSF Award \#1453386 and AFRL Award \#8750-14-2-0270. The views and conclusions contained herein are those of the authors and should not be interpreted as necessarily representing the official policies or endorsements, either expressed or implied, of the U.S. Government. 
\end{acks}

%% file: appendix.tex
\appendix

\section{Correctness Proof of CFTA Method}

\section{Proofs of Theorems}

\vspace{0.1in}
\noindent
{\sc Theorem 3.3}
{\bf (Soundness of AFTA)}
Let $\fta$ be the AFTA constructed for examples $\exs$ and grammar $\grammar$ using the abstraction defined by finite set of predicates $\mathcal{P}$ (including $\emph{true}$). If $\prog$ is a program that is consistent with examples $\exs$, then $\prog$ is accepted by $\fta$. 

\begin{proof}
Suppose program $\prog$ is represented by its AST with nodes $V$ and $\fta$ is given as $\fta = (\ftastates, \alphabet, \finalstates, \transitions)$. Furthermore, we assume $\prog_{v}$ denotes the sub-AST rooted at node $v \in V$. 
In what follows, we prove by structural induction that using rewrite rules in $\transitions$ any node $v \in V$ is rewritten to a state $\ftastate_{s}^{\vec{\aval}}$ such that $\forall j \in [1, |\vec{\ex}|]. \ (s = \semantics{\prog_{v}} \ex_{\inp, j}) \refines \aval_{j}$, i.e., $\aval_j$ overapproximates $\semantics{\prog_{v}} \ex_{\inp, j}$.  
\begin{enumerate}
\item 
Base case: $v$ is a leaf node, i.e., $x$ or constant. 
According to Var and Const rules from \figref{abstractrules}, we have $\aval_j = \alpha^{\preds}(s = \semantics{\prog_{v}} \ex_{\inp, j})$ where $s$ is the label of $v$. Thus, we have $(s = \semantics{\prog_{v}} \ex_{\inp, j}) \refines \aval_j$. 
\item 
Inductive case: $v$ is a non-leaf node. 
Suppose the children ASTs of $v$ are given as $\vec{\prog}$, and the label of $v$ is $f$. 
According to the inductive hypothesis, we have $\forall i \in [1, |\vec{\prog}|].$ \emph{root}($\prog_i$) is rewritten to a state $\ftastate_{s_i}^{\vec{\aval_i}}$ such that $(s_i = \semantics{\prog_{i}} \ex_{\inp, j}) \refines \aval_{ij}$. According to Prod rule from \figref{abstractrules}, we have $\aval_j = \alpha^{\preds} \big{(} \asemantics{f(\aval_{1j}, \dots, \aval_{|\vec{\prog}| j })} \big{)}$. In another word, we have $\asemantics{f(\aval_{1j}, \dots, \aval_{|\vec{\prog}| j })} \refines \aval_j$. Since our abstract transformers are sound, we have $(s = \semantics{\prog_{v}} \ex_{\inp, j}) \refines \asemantics{f(\aval_{1j}, \dots, \aval_{|\vec{\prog}| j })}$. Therefore, we have $(s = \semantics{\prog_{v}} \ex_{\inp, j}) \refines \aval_j$. 
\end{enumerate}
Now, we have proved that using transitions in $\transitions$ any node $v \in V$ is rewritten to a state $\ftastate_{s}^{\vec{\aval}}$ such that $\forall j \in [1, |\vec{\ex}|]. \ (s = \semantics{\prog_{v}} \ex_{\inp, j}) \refines \aval_{j}$. 
As a special case, we have that the root node $\emph{root}(\prog)$ is rewritten to a state $\ftastate_{s_0}^{\vec{\aval}}$ such that $(s_0 = \semantics{\prog} \ex_{\inp, j}) \refines \aval_j \ (j \in [1,|\vec{\ex}|])$. Since $\prog$ is consistent with examples $\vec{\ex}$, we have $\semantics{\prog} \vec{\ex}_{\inp} = \vec{\ex}_{\out}$, and thus, $(s_0 = \ex_{\out, j}) \refines \aval_j \ (j \in [1,|\vec{\ex}|])$. According to Final rule from \figref{abstractrules}, $\ftastate_{s_0}^{\vec{\aval}}$ is a final state. Therefore, $\prog$ is accepted by $\fta$. 
\end{proof}

\noindent
{\sc Theorem 4.2}
{\bf (Existence of Proof)}
Given a spurious program $\prog$ that does not satisfy example $\ex$, we can always find a proof of incorrectness of $\prog$ satisfying the properties from Definition~\ref{def:itp}.

\begin{proof}
Suppose $\prog$ is represented by its AST with nodes $V$, and $\prog_{v_i}$ denotes the sub-AST rooted at node $v_i \in V$. 
Let $c_i$ be $\semantics{\prog_{v_i}} \ex_{\inp} \ (v_i \in V)$. Now, let us consider the proof $\itp^\prime$ that annotates each node $v_i \in V$ with the predicate $s_i = c_i$ where $s_i$ is the grammar symbol of $v_i$. It is obvious that $\itp^\prime$ is a proof that satisfies properties (1) and (2) from Definition~\ref{def:itp}, since essentially $\itp^{\prime}$ corresponds each AST node to its concrete value on $\ex_{\inp}$. Furthermore, since program $\prog$ does not satisfy example $\ex$, i.e., $\semantics{\prog} \ex_{\inp} \neq \ex_{\out}$, we have $\ex_{\out} \not\in \gamma(s_0 = \semantics{\prog} \ex_{\inp})$. Thus, property (3) from Definition~\ref{def:itp} also holds. Therefore, $\itp^{\prime}$ is a proof of incorrectness of $\prog$ satisfying all three properties from Definition~\ref{def:itp}. Because we assume that the universe $\universe$ always includes the predicates of the form $s = c$ for any grammar symbol $s$ and any concrete value $c$ that symbol $s$ can take, therefore, we can always find a proof of incorrectness of $\prog$ satisfying the properties from Definition~\ref{def:itp}. 
\end{proof}

\noindent
{\sc Theorem 4.3}
{\bf (Progress)}
Let $\fta_i$ be the AFTA constructed during the i'th iteration of the {\sc Learn} algorithm from \figref{synthesisalgorithm}, and let $\prog_i$ be a spurious program returned by {\sc Rank}, i.e., $\prog_i$ is accepted by $\fta_i$ and does not satisfy input-output examples $\ex$. Then, we have $\prog_i \not \in \mathcal{L}(\fta_{i+1})$ and $\mathcal{L}(\fta_{i+1}) \subset \mathcal{L}(\fta_{i})$.

\begin{proof}
We first prove that $\prog_i \not \in \mathcal{L}(\fta_{i+1})$. Since $\semantics{\prog_i} \ex_{\inp} \neq \ex_{\out}$, there exists a proof of incorrectness that establishes the spuriousness of $\prog_i$ according to {\sc Theorem 4.2}. Assume the {\sc ConstructProof} procedure finds a proof $\itp_i$, i.e., $\prog$ can be shown to be incorrect using the new predicates in $\itp_i$ as well as predicates in $\preds_i$ (the set of predicates used in the i'th iteration). In (i+1)'th iteration, we have $\preds_{i+1} = \preds_i \cup \textsc{ExtractPredicates}(\itp_i)$ according to the {\sc Learn} procedure from \figref{synthesisalgorithm}. Combined with construction rules from \figref{abstractrules}, we know that $\emph{root}(\prog)$ will be rewritten to a state $\ftastate_{\outputsymbol}^{\vec{\aval}}$ such that $(s_0 = \ex_{\out}) \not\refines \aval_j$ ($j$ is the index of $\ex$ in $\vec{\ex}$). Therefore, $\prog_i$ is not accepted by $\fta_{i+1}$ (according to Final rule from \figref{abstractrules}). 

Now we prove that $\mathcal{L}(\fta_{i+1}) \subset \mathcal{L}(\fta_{i})$. We first show that $\mathcal{L}(\fta_{i+1}) \subseteq \mathcal{L}(\fta_{i})$. This is obvious since any program which is accepted by $\fta_{i+1}$ should also accepted by $\fta_i$ (recall that $\fta_{i+1}$ is constructed using predicates $\preds_{i+1} \supseteq \preds_i$). Furthermore, since we proved that we have $\prog_i \not\in \mathcal{L}(\fta_{i+1})$ for the program $\prog_i \in \mathcal{L}(\fta_i)$, we have $\mathcal{L}(\fta_{i+1}) \subset \mathcal{L}(\fta_{i})$. 

Therefore, this theorem holds. 
\end{proof}

\noindent
{\sc Theorem 4.5}
{\bf (Soundness and Completeness)}
If there exists a DSL program that satisfies the input-output examples $\exs$, then the {\sc Learn} procedure from \figref{synthesisalgorithm} will return a program $\prog$ such that $\semantics{\prog} \exs_\inp = \exs_\out$.

\begin{proof}
Suppose the DSL program that satisfies $\exs$ is $\prog^{\prime}$. Let the cost of $\prog^{\prime}$ given by the {\sc Rank} function be $C'$. Since {\sc Rank} defines a deterministic order of programs according to their costs, we know that there are finitely many programs all of which have costs no greater than $C'$. Let us use $S$ to denote these programs (apparently we have $\prog' \in S$). We know that there exists at least one program in $S$ (i.e., $\prog'$) that satisfies the examples $\exs$. 
Since in each iteration, our synthesis procedure {\sc Learn} finds the program with the minimum cost in the current search space (defined by the language of the AFTA), it takes at most $|S|$ iterations to find a program $\prog$ such that $\semantics{\prog} \exs_\inp = \exs_\out$. 
\end{proof}

\noindent
{\sc Theorem 4.7}
{\bf (Correctness of Proof)}
The mapping $\itp$ returned by the {\sc ConstructProof} procedure satisfies the properties from Definition~\ref{def:itp}.

\begin{proof}
$\itp$ satisfies property (3) from Definition~\ref{def:itp} because the {\sc StrengthenRoot} procedure (lines 5-7) from \figref{strengthenroot} is guaranteed to find an annotation for the root node such that property (3) is satisfied.  
Furthermore, $\itp$ also satisfies properties (1) and (2) from Definition~\ref{def:itp} because the {\sc StrengthenChildren} procedure (lines 6-8) is guaranteed to find annotations for all children of any AST node such that properties (1) and (2) are established. Therefore, we conclude the proof. 
\end{proof}